\documentclass[preprint,review,12pt]{elsarticle}

\usepackage{subcaption}
\usepackage{amssymb}
\usepackage{amsmath}
\usepackage{booktabs}
\usepackage{graphicx}
\usepackage{hyperref}

\journal{Renewable Energy}

\begin{document}

\begin{frontmatter}

\title{Adaptive optimization of wave energy conversion in oscillatory wave surge converters via SPH simulation and deep reinforcement learning}

\author[inst1]{Mai Ye}

\author[inst2]{Chi Zhang}

\author[inst3]{Yaru Ren}

\author[inst2]{Ziyuan Liu}

\author[inst1]{Oskar J. Haidn}

\author[inst1]{Xiangyu Hu\texorpdfstring{\corref{cor1}}{}}
\ead{xiangyu.hu@tum.de}

\cortext[cor1]{Corresponding author}

\affiliation[inst1]{organization={TUM School of Engineering and Design, Technical University of Munich},
            postcode={85748}, 
            city={Garching},
            country={Germany}}

\affiliation[inst2]{organization={Huawei Technologies Munich Research Center}, 
            postcode={80992}, 
            city={Munich},
            country={Germany}}

\affiliation[inst3]{organization={State Key Laboratory of Hydraulics and Mountain River Engineering, Sichuan University}, 
            postcode={610065}, 
            city={Chengdu},
            country={China}}

\begin{abstract}
The nonlinear damping characteristics of the oscillating wave surge converter (OWSC) significantly impact the performance of the power take-off system. This study presents a framework by integrating deep reinforcement learning (DRL) with numerical simulations of OWSC to identify optimal adaptive damping policy under varying wave conditions, thereby enhancing wave energy harvesting efficiency. Firstly, the open-source multiphysics libraries SPHinXsys and Simbody are employed to establish the numerical environment for wave interaction with OWSCs. Subsequently, a comparative analysis of three DRL algorithms—proximal policy optimization (PPO), twin delayed deep deterministic policy gradient (TD3), and soft actor-critic (SAC)—is conducted using the two-dimensional (2D) numerical study of OWSC interacting with regular waves. The results reveal that artificial neural networks capture the nonlinear characteristics of wave-structure interactions and provide efficient PTO policies. Notably, the SAC algorithm demonstrates exceptional robustness and accuracy, achieving a 10.61\% improvement in wave energy harvesting. Furthermore, policies trained in a 2D environment are successfully applied to the three-dimensional (3D) study, with an improvement of 22.54\% in energy harvesting. Additionally, the study shows that energy harvesting is improved by 6.42\% for complex irregular waves. However, for the complex dual OWSC system, optimizing the damping characteristics alone is insufficient to enhance energy harvesting.
\end{abstract}


\end{frontmatter}

\section{Introduction}
Considering the significant environmental issues caused by the extensive use of fossil fuels, including pollution, greenhouse gas emissions, and ecological destruction, there has been a marked increase in the study of clean and renewable energy sources. Wind and solar power have seen rapid technological advancements over the past few decades, achieving successful commercial applications. However, as energy demand grows, a broader range of renewable energy sources has come into focus. Among these, wave energy stands out due to its substantial potential (with a minimum estimated capacity of around 0.2 TW), high energy density, strong stability, and the advantage of not occupying land resources \cite{callaway2007catch}. These features have attracted significant research and development investments, making wave energy a crucial component in transitioning to a sustainable energy future. Historically, most research has focused on extracting energy from the heave motion of deep-water systems, mainly due to the common belief that nearshore wave resources are significantly lower than those in deeper waters \cite{budal1982wave, evans1981maximum}. However, since the beginning of the 21st century, the concept of exploitable wave energy resources has become more realistic \cite{folley2009analysis}. In many nearshore locations, the exploitable resource is typically only 10-20\% lower than that offshore \cite{folley2010analysis}. As a result, the extraction of wave energy from the surge motion of waves in nearshore waters has gained increasing attention.

\begin{figure}[htbp]
\centering
\includegraphics{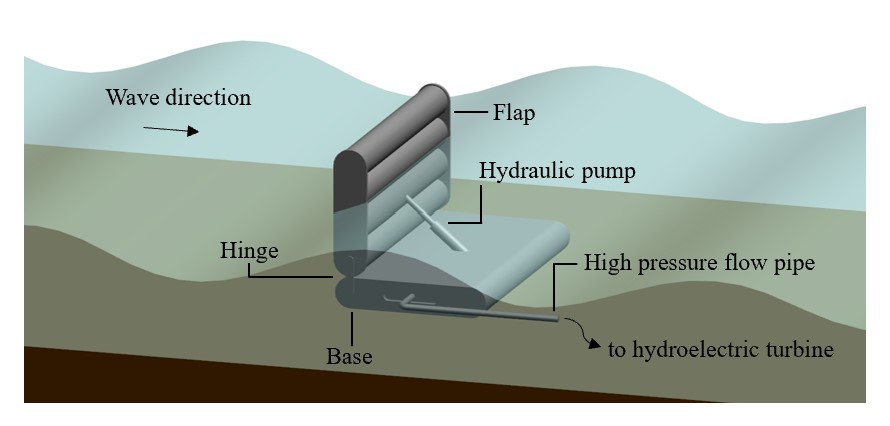}
\caption{Schematic of Oyster\textsuperscript{\textregistered} (a type of OWSC) under waves.}\label{fig1}
\end{figure}

Typical wave energy converters (WECs) can be classified into three main categories based on their working principles: oscillating water column (OWC) devices, which use the oscillating water column to compress air and drive a turbine \cite{lopez2019site}, over-topping devices, which utilize the potential energy of waves as they spill over a barrier \cite{kofoed2006prototype}, and wave-activated bodies, which exploit the heave, surge, roll, or pitch motions depending on their construction \cite{ilyas2014wave, day2015hydrodynamic}. oscillating wave surge converters (OWSCs) are typical wave-activated bodies used in nearshore waters, usually employing bottom-hinged flap mechanisms. Notable examples include the products of WaveRoller and Oyster \cite{whittaker2012nearshore, folley2007design}. The primary distinction between these two lies in the positioning of their flaps: WaveRoller's flap is wholly submerged in seawater, whereas Oyster's flap has an upper edge that protrudes above the water surface \cite{cheng2019fully}. Research conducted at Queen's University Belfast suggests that while partial submersion enhances the impact pressure exerted by the rotating flap, the flap inherently decouples from the wave as the oscillation amplitude increases \cite{whittaker2007development}. This decoupling effect ensures that the wave-induced loads remain manageable, thereby safeguarding the structural integrity of the Oyster, even under extreme sea conditions. The structure of the Oyster is shown in Fig.~\ref{fig1}. The flap is connected to the base via a hinge and oscillates back and forth in response to the incident waves. This oscillatory motion drives the power take-off (PTO) system, which utilizes a hydraulic pump to channel high-pressure water through a pipeline to a hydroelectric turbine, generating electricity \cite{renzi2014does}.

According to the experimental study of the wave interacting with OWSCs by Henry et al. \cite{henry2013characteristics}, the flap of an OWSC actively impacts the trough of incoming waves, generating breaking waves. Chow et al. \cite{chow2018experimental} discovered that a system of two tandem OWSCs could improve overall wave energy conversion efficiency and determined the optimal separation distance between them based on Bragg reflection. Brito et al. \cite{brito2020experimental} demonstrated through scale model experiments that the capture width ratio (CWR) and response amplitude operator (RAO) of an OWSC exhibit weak correlation under both regular and irregular wave conditions. 

Numerical study is attracting more and more attention as it can be applied to to explore the detailed mechanisms of wave-structure interactions (WSI), optimize the structure of OWSCs, and improve the energy harvesting efficiency of PTO systems. For example, Folley et al. \cite{folley2007design} utilized a linearized frequency domain model based on the Boundary Element Method (BEM). As a following, Cheng et al. \cite{cheng2019fully, cheng2020fully} developed a two-dimensional (2D) and three-dimensional (3D) higher-order boundary element method (HOBEM) model to estimate the performance of an OWSC. Renzi and Dias \cite{renzi2012resonant} also proposed a semi-analytical model for 3D computation of OWSCs. Although these methods offer high computational efficiency, they lack the accuracy of predicting nonlinear phenomena such as slamming and overtopping. Two approaches for this problem are based on the Navier-Stokes equations with either mesh-based methods or particle-based methods. Mesh-based methods are primarily implemented using commercial software such as Ansys Fluent and the open-source platform OpenFOAM. Schmitt et al. were the first to use Reynolds-averaged Navier–Stokes (RANS) equations to explore the nonlinear relationship between wave height and the optimal damping of the OWSC \cite{schmitt2016optimising}. The numerical simulations by Wei et al. \cite{wei2015wave, wei2016wave} also agreed with experimental results in predicting the wave height and the pressure distribution on the flap. They found that the effects of viscosity and Froude scaling on OWSCs are minimal, whereas re-reflection significantly enhances or suppresses the incident waves and impacts. Jiang et al. \cite{jiang2018hydrodynamic} discussed the impact of different damping strategies on power generation efficiency in PTO systems. They found that nonlinear damping strategies can enhance system stability but do not significantly increase power output. However, these methods come with significant computational costs due to the inclusion of additional Volume of Fluid (VOF) equations and the necessity of using dynamic mesh techniques to solve for the motion of the flap \cite{schmitt2015use, mottahedi2018application}. 

Particle-based methods, i.e., SPH, are particularly well-suited for addressing challenges involving large deformations, complex free surface flows, and multiphase flows \cite{luo2017shared, zhang2017generalized, khayyer2018development}, making them a compromising alternative to study the hydrodynamic interactions between waves and WECs. Specifically, for OWSC calculations, Henry et al. \cite{henry2013characteristics} and Rafiee et al. \cite{rafiee2013numerical} used a modified SPH method to perform 2D and 3D numerical simulations. They employed a Riemann solver to compute wave density and incorporated the Lagrangian form of the RANS $k-\epsilon$ model into the equations to capture the turbulence characteristics. The results indicated that 3D simulations can more accurately predict the pressure distribution on the flap. Brito et al. \cite{brito2016coupling} presented a numerical model combining DualSPHysics for wave computation and Chrono for nonlinear mechanical constraint systems of OWSCs. Later, they updated the numerical model to consider most constraints, such as the PTO system, revolute joints, and frictional contacts \cite{brito2020numerical}. Wei et al. \cite{wei2019modeling} proposed a similar solution, replacing the SPH library with the GPUSPH code. Their simulation results showed that when the natural frequency of the OWSC structure is similar to the wave period, it can maximize power capture. Liu et al. \cite{liu2020numerical} quantitatively analyzed the effects of parameters such as load, flap mass, thickness, hinge height, and damping of the PTO system on motion resonance and wave absorption of OWSC. Zhang et al. \cite{zhang2021efficient} used a Riemann-based weakly compressible method based on SPHinXsys to compute incident waves and WSI. Simobdy is employed for solid dynamics. Their results demonstrated that this solver could accurately predict wave height and the pressure distribution on the flap while significantly reducing computation time, showing great potential for practical applications.

Currently, the performance optimization of OWSCs primarily relies on experiments, theoretical analysis, and numerical simulations under specific wave conditions to conduct parametric studies on the structure and position of flaps or damping of the PTO system \cite{gomes2015dynamics, senol2019enhancing, liu2022performance, liu2022nonlinear}. There is still a lack of research on effective control strategies to enhance the performance of OWSCs. In comparison, control strategy methods have already been applied to optimize the wave energy absorption performance of other WEC devices \cite{ringwood2014energy, jia2020economic, shadman2021power}. However, the interaction between waves, especially irregular waves, and OWSCs is complex, highly stochastic, and nonlinear. Models such as latching and model predictive control (MPC) exhibit poor robustness and are highly dependent on the accuracy of the predictive model. Inaccurate models can significantly affect performance \cite{liang2023energy}. In the past decade, with the rapid advancement of artificial intelligence (AI), deep reinforcement learning (DRL) has demonstrated significant capabilities in the field of active flow control \cite{fan2019robotic, fan2020reinforcement, rabault2020deep, garnier2021review}. DRL combines artificial neural networks (ANN) and reinforcement learning (RL). Since ANN uses nonlinear activation functions and can effectively fit any function, DRL, compared to traditional RL, enhances the exploration and capture of high-dimensional state spaces, making it suitable for WSI problems \cite{hornik1989multilayer}. Rabault et al. \cite{rabault2019artificial} were the first to use DRL to achieve drag reduction in the problem of flow around a cylinder through real-time control of the jets on both sides. Research on optimizing WECs using DRL employs potential wave theory as the environment. Studies have shown that DRL can enhance the energy harvesting efficiency of WECs \cite{anderlini2020towards, zou2022optimization}. Additionally, compared to directly using computational fluid dynamics (CFD) as the environment, this approach significantly reduces the time required to train the ANN. However, potential wave theory is less effective at capturing the coupling effects between multiple physical fields. Only Liang et al. \cite{liang2024environmental} have combined 2D CFD with DRL to optimize the wave energy conversion of horizontal floating cylinders under irregular waves.

In this paper, we will first establish a new platform that combines CFD with DRL. Based on our previous work, we will use SPHinXsys, a multi-physics library based on the SPH method, and Simbody as the numerical computation platform for the bottom-hinged OWSC \cite{zhang2021sphinxsys}. Given that mainstream DRL platforms like Tianshou employ neural networks such as PyTorch, which are based on the Python environment, we will use Pybind11 to package the relevant OWSC code into a dynamic link library for invocation in the OpenAI Gymnasium environment \cite{weng2022tianshou}. This standardized environment will facilitate the direct application of various DRL algorithms available on the Tianshou platform, enabling us to explore the impact of these algorithms on performance improvement. Additionally, in the specific training process, we will not simply define the damping coefficient of the PTO system as linear, rather, it will be a parameter that varies in real-time with the incident waves and the motion state of the flap. This parameter will be controlled using a neural network optimized by the RL algorithm. 

The remainder of this paper is organized as follows: Section 2 introduces the Riemann-based SPH method for hydrodynamic modeling in SPHinXsys, the simulation of dynamic damping characteristics using Simbody, the detailed coupling process, wave-making theory, and validation against theoretical and experimental results. Section 3 discusses the details of our CFD-DRL framework, the DRL training environment, and the mainstream RL algorithms. Section 4 analyzes the performance differences of various DRL algorithms in the OWSC problem, explores the applicability of 2D and 3D numerical computations to the policies, verifies that the adaptive damping coefficient policy can be applied to random irregular waves, and investigates the feasibility of wave energy optimization under the dual OWSC systems comparing it with the optimal layout and the best combination of constant damping coefficients.

\section{Numerical modeling}
Accurate numerical simulation of the wave interaction with OWSCs is crucial for building a reliable environment for RL. In this section, we utilize the Lagrange-based numerical computing library SPHinXsys to perform precise calculations. Specifically, we introduce an improved weakly compressible SPH (WCSPH) method \cite{zhang2017weakly}, corrected with weighted kernel gradient (WKG) \cite{ren2023efficient} to handle the accurate prediction of wave dynamic and its interaction with the OWSC. Furthermore, we present a comprehensive coupling approach between SPHinXsys and the Simbody library to capture the WSI.

\subsection{Governing equation}
In the Lagrangian framework, the mass and momentum conservation equations for incompressible and viscous fluid can be written as
\begin{equation}
\left\{
\begin{aligned}
\label{eq:eq1}
\frac{d\rho}{dt} &= -\rho\nabla \cdot \mathbf{v} \\
\frac{d\mathbf{v}}{dt} &= -\frac{1}{\rho}\nabla p + \nu\nabla^2\mathbf{v} + \mathbf{g}.
\end{aligned}
\right.
\end{equation}
Here, $\rho$ is the density of the fluid, $\mathbf{v}$ the velocity, $p$ the pressure, $\nu$ the kinematic viscosity, and $\mathbf{g}$ is the gravity. An artificial isothermal equation of state (EoS) is used to close the system of Eq.~(\ref{eq:eq1})
\begin{equation}
\label{eq:eq2}
p = c^2(\rho - \rho^{0}),
\end{equation}
where c is the artificial speed of sound, $\rho^{0}$ the initial density. Following the weakly-compressive assumption, $c = 10v_{max}$ is applied to ensure that the density varies around 1\% \cite{morris1997modeling}. Here, $v_{max}$ is the maximum anticipated particle velocity in the flow. In this paper, we set $v_{max} = 2\sqrt{gh}$ with $h$ denoting the water depth.

\subsection{Riemann-based WCSPH method}
In this work, Eq.~(\ref{eq:eq1}) is discretized with Riemann-SPH method \cite{zhang2017weakly} as 
\begin{equation}
\left\{
\begin{aligned}
\label{eq:eq3}
\frac{d\rho_{i}}{dt} &= 2\rho_{i}\sum_{j}V_{j}(U^{*} - \mathbf{v}_{i} \cdot \mathbf{e}_{ij})\frac{\partial W_{ij}}{\partial r_{ij}} \\
\frac{d\mathbf{v}_{i}}{dt} &= -m_{i}\sum_{j}\frac{2P^*}{\rho_{i}\rho_{j}}\nabla_{i}W_{ij} +  m_{i}\sum_{j}\frac{2\mu}{\rho_{i}\rho_{j}}\frac{\mathbf{v}_{ij}}{r_{ij}}\frac{\partial W_{ij}}{\partial r_{ij}} + \mathbf{g}_i.
\end{aligned}
\right.
\end{equation}
Here, $m_{i}$ and $\rho_{i}$ are the mass and density of particle $i$, $V_{j}$ the particle volume, $\mathbf{v}_{ij} = \mathbf{v}_{i} - \mathbf{v}_{j}$ particle relative velocity, and $\mu$ is the dynamic viscosity. Also, $\nabla W_{ij} =\mathbf{e}_{ij} ( \partial W({r}_{ij}, h) / \partial r_{ij})$ with $\mathbf{e}_{ij} = \mathbf{r}_{ij} / r_{ij}$ and $\mathbf{r}_{ij} = \mathbf{r}_{i} - \mathbf{r}_{j}$, and ${W}_{ij}$ represents the Kernel gradient \cite{wendland1995piecewise}. Besides, $U^{*}$ and $P^*$ are the solutions of the one-dimensional Riemann problem constructed along the line pointing from particle $i$ to $j$. The left and right initial states of the Riemann problem can be reconstructed as
\begin{equation}
\left\{
\begin{aligned}
\label{eq:eq4}
(\rho_{L}, U_{L}, P_{L}, c_{L}) &= (\rho_{i}, -\mathbf{v}_{i} \cdot \mathbf{e}_{ij}, p_{i}, c_{i}) \\
(\rho_{R}, U_{R}, P_{R}, c_{R}) &= (\rho_{j}, -\mathbf{v}_{j} \cdot \mathbf{e}_{ij}, p_{j}, c_{j}).
\end{aligned}
\right.
\end{equation}

To solve this one-dimensional Riemann problem, Zhang et al. \cite{zhang2017weakly} adopted a linearised Riemann solver coupled with a dissipation limiter
\begin{equation}
\left\{
\begin{aligned}
\label{eq:eq5}
U^{*} &=  \frac{\rho_{L}c_{L}U_{L} + \rho_{R}c_{R}U_{R} + P_{L} - P_{R}}{\rho_{L}c_{L} + \rho_{R}c_{R}} \\
P^{*} &=  \frac{\rho_{L}c_{L}P_{R} + \rho_{R}c_{R}P_{L} + \rho_{L}c_{L}\rho_{R}c_{R}\beta(U_{L} - U_{R})}{\rho_{L}c_{L} + \rho_{R}c_{R}},
\end{aligned}
\right.
\end{equation}
with low dissipation limiter $\beta = min(3max(U_{L} - U_{R}, 0) / \overline{c}, 1)$ where $\overline{c} = (\rho_{L}c_{L} + \rho_{R}c_{R}) / (\rho_{L} + \rho_{R})$.

\subsection{Weighted kernel gradient correction}
Current research indicates that using the Riemann-based WCSPH method to solve the wave dynamics leads to significant numerical dissipation, which affects the simulation of small waves \cite{kanehira2022effects}. The kernel gradient correction (KGC) method can effectively reduce the numerical dissipation and improve energy conservation \cite{wen2018improved}. 

Following Ref. \cite{monaghan1992smoothed}, the correction matrix in Eq.~(\ref{eq:eq6}) is introduced to modify the kernel gradient calculation to ensure first-order consistency.
\begin{equation}
\label{eq:eq6}
\mathbf{B}_{i} = (-\sum_{j}\mathbf{r}_{ij} \otimes \nabla_{i}{W}_{ij}V_{j})^{-1} = (\mathbf{A}_{i})^{-1}
\end{equation}

However, this correction encounters problems such as computational instability when the determinant of the matrix $\mathbf{A}_{i}$ approaches zero. In order to improve the robustness of numerical simulation, a generalized and consistent weighted kernel gradient correction (WKGC) proposed by Ren et al. \cite{ren2023efficient} is adopted as
\begin{equation}
\left\{
\begin{aligned}
\label{eq:eq7}
\widetilde{\mathbf{B}_{i}} &= \omega_{1}\mathbf{B}_{i} + (1-\omega_{1})\mathbf{I} \\
\omega_{1} &= \frac{\left|\mathbf{A}_{i}\right|}{\epsilon + \left|\mathbf{A}_{i}\right|}.
\end{aligned}
\right.
\end{equation}
Here, $\mathbf{I}$ is the identity matrix, and $\epsilon = 0.3$.

For the Riemann-based WCSPH, the kernel correction matrix is introduced to modify the solution of pressure item $P^{*}$ in Eq.~(\ref{eq:eq5}) as
\begin{equation}
\label{eq:eq8}
P^{*} =  \frac{\rho_{L}c_{L}P_{R}\widetilde{\mathbf{B}_{i}} + \rho_{R}c_{R}P_{L}\widetilde{\mathbf{B}_{j}} + \rho_{L}c_{L}\rho_{R}c_{R}\beta(U_{L} - U_{R})}{\rho_{L}c_{L} + \rho_{R}c_{R}}.
\end{equation}

\subsection{Dual-criteria time stepping}
In order to improve computational efficiency, dual-criteria time-stepping method is adopted here. Specifically, the update frequency of the particle configuration is controlled by the advection criterion, while the integration of pressure relaxation is determined by a smaller time step size based on the acoustic criterion. Following Zhang et al. \cite{zhang2017weakly}, the time step size of the advection criterion $\Delta t_{ad}$ and the acoustic criterion $\Delta t_{ac}$ are
\begin{equation}
\left\{
\begin{aligned}
\label{eq:eq9}
\Delta t_{ad} &= CFL_{ad}\min(\frac{h}{|\mathbf{v}|_{max}}, \frac{h^2}{\nu}) \\
\Delta t_{ac} &= CFL_{ac}(\frac{h}{c + |\mathbf{v}|_{max}}),
\end{aligned}
\right.
\end{equation}
where $CFL_{ad} = 0.25$ and $CFL_{ac} = 0.6$. More details are refered to Ref. \cite{zhang2017weakly}.

\subsection{SPHinXsys and Simbody coupling}
To compute the solid-body kinematics in wace interaction with OWSCs, we adopted the Simbody library, which provides an object-oriented C++ API. This feature allows it to be coupled with SPHinXsys and exposed to the Python environment. In the current framework, all media, including rigid bodies such as the flap and hinge, are presented with particles. The forces exerted from the fluid on the rigid bodies and damping coefficients are transmitted to Simbody via SPHinXsys. For predicting the combined translational and rotational motion by solving the Newton-Euler equations.

In detail, the entire calculation sequence begins with the calculation of the fluid properties. The fluid density $\rho_{i}$ will be initialized firstly at the beginning of the advection time step $\Delta t_{ad}$ as
\begin{equation}
\label{eq:eq10}
\rho_{i} = \max(\rho^{*}, \rho^{0}\frac{\sum W_{ij}}{\sum W_{ij}^{0}}),
\end{equation}
where $\rho^{*}$ denotes the density before re-initialization and $\rho^{0}$ represents the initial reference value. The viscous force $\mathbf{f}_{a\nu}$ is also computed at this stage. Subsequently, pressure relaxation is carried out over the next several acoustic time steps $\Delta t_{ac}$ using the position-based Verlet scheme proposed by Zhang et al. \cite{zhang2020dual}
\begin{equation}
\left\{
\begin{aligned}
\label{eq:eq11}
\rho_{i}^{n + \frac{1}{2}} &= \rho_{i}^{n} + \frac{1}{2}\Delta t_{ac}(\frac{d \rho_{i}}{d t})^{n + \frac{1}{2}} \\
\mathbf{r}_{i}^{n + \frac{1}{2}} &= \mathbf{r}_{i}^{n} + \frac{1}{2}\Delta t_{ac}\mathbf{v}_{i}^{n}.
\end{aligned}
\right.
\end{equation}

After that the particle's velocity $\mathbf{v}_{i}$, density $\rho_{i}$, and position $\mathbf{r}_{i}$ are updated to the mid-points as
\begin{equation}
\left\{
\begin{aligned}
\label{eq:eq12}
\mathbf{v}_{i}^{n + 1} &= \mathbf{v}_{i}^{n} + \Delta t_{ac}(\frac{d \mathbf{v}_{i}}{d t})^{n + 1} \\
\rho_{i}^{n + 1} &= \rho_{i}^{n + \frac{1}{2}} + \frac{1}{2}\Delta t_{ac}(\frac{d \rho_{i}}{d t})^{n + \frac{1}{2}} \\
\mathbf{r}_{i}^{n + 1} &= \mathbf{r}_{i}^{n + \frac{1}{2}} + \frac{1}{2}\Delta t_{ac}\mathbf{v}_{i}^{n + 1}.
\end{aligned}
\right.
\end{equation}

Then, the forces acting on the flap of the OWSC will be computed, which are composed of two main components, the pressure force $\mathbf{f}_{ap}$ and the viscous force $\mathbf{f}_{a\nu}$, as
\begin{equation}
\left\{
\begin{aligned}
\label{eq:eq13}
\mathbf{f}_{ap} &= -2\sum_{i}V_{i}V_{a}\frac{p_{i}\rho_{a}^{d} + p_{a}^{d}\rho_{i}}{\rho_{i} + \rho_{a}^{d}}\nabla_{a}W_{ai} \\
\mathbf{f}_{a\nu} &= 2\sum_{i}\nu V_{i}V_{a}\frac{\mathbf{v_{i} - \mathbf{v_{a}^{d}}}}{r_{ai}}\frac{\partial W_{ai}}{\partial r_{ai}}.
\end{aligned}
\right.
\end{equation}
Here, the subscript $a$ denotes the solid particle index. The imaginary pressure $p_{a}^{d}$ and velocity $\mathbf{v}_{a}^{d}$ read as follows
\begin{equation}
\left\{
\begin{aligned}
\label{eq:eq14}
p_{a}^{d} &= p_{i} + \rho_{i}\max{(0, (\mathbf{g} - \frac{\mathbf{dv}_{a}}{dt}) \cdot \mathbf{n})(\mathbf{r}_{ai} \cdot \mathbf{n})} \\
\mathbf{v}_{a}^{d} &= 2\mathbf{v}_{i} - \mathbf{v}_{a},
\end{aligned}
\right.
\end{equation}
where $\mathbf{n}$ represents the normal direction of the solid body to fluid. The total force and total torque $\tau$ acting on the flap can be written as
\begin{equation}
\left\{
\begin{aligned}
\label{eq:eq15}
\mathbf{F} &= \sum_{a\in N}\mathbf{f}_{a} = \sum_{a\in N}(\mathbf{f}_{ap} + \mathbf{f}_{a\nu}) \\
\tau &= \sum_{a\in N}(\mathbf{r}_{a} - \mathbf{r}_{G}) \times \mathbf{f}_{a},
\end{aligned}
\right.
\end{equation}
where $N$ denotes the total number of solid particles and $\mathbf{r}_{G}$ is the position vector of the flap mass center. 

At the end of each time step, the total force and total torque calculated from SPHinXsys are transmitted to Simbody for solving the Newton-Euler equations
\begin{equation}
\left\{
\begin{aligned}
\label{eq:eq16}
\mathbf{F} &= m\mathbf{I}_{0}\frac{d\mathbf{v}}{dt} \\
\tau &= \mathbf{J}_{0}\frac{d\Omega}{dt} - k_{d}\Omega.
\end{aligned}
\right.
\end{equation}
where $m$ is the mass of the flap, $\mathbf{I}_{0}$ the identity matrix, $\mathbf{J}_{0}$ the moment of the inertia about the center of mass, $\omega$ the angular velocity and $k_{d}$ the damping coefficient. After updating the position, velocity, and normal direction, the new kinematic state will be imported back to SPHinXsys, and the loop will continue.

It is worth noting that previous research \cite{zhang2021efficient} introduced a linear damper for modeling the PTO system of the OWSC. The damping coefficient $k_{d}$ was fixed during the simulation. In this paper, during the computation of Eq.~(\ref{eq:eq16}), the $k_{d}$ from the policy is variable within a specific range, and it can be rapidly implemented through the damping update definition in Simbody.

\subsection{Wave generation}
In the Riemann-based WCSPH method, the solid wall boundary can be discretized by dummy particles. For fluid particles around the wall region, the interaction is determined by solving a one-sided Riemann problem along the wall-normal direction \cite{zhang2017weakly, adami2012generalized} where the left state is defined
\begin{equation}
\label{eq:eq17}
(\rho_{L}, U_{L}, P_{L}) = (\rho_{i}, -\mathbf{n}_{w} \cdot \mathbf{v}_{i}, p_{i}),
\end{equation}
where $\mathbf{n}_{w}$ is the normal direction of the wall, and $i$ represents the fluid particles. The right state of the velocity $U_{R}$ and pressure $P_{R}$ are assumed as
\begin{equation}
\left\{
\begin{aligned}
\label{eq:eq18}
U_{R} &= - U_{L} + 2\mathbf{u}_{w} \\
P_{R} &= P_{L} + \rho_{i}\mathbf{g} \cdot \mathbf{r}_{iw},
\end{aligned}
\right.
\end{equation}
where $\mathbf{u}_{w}$ is wall velocity, $\mathbf{r}_{iw} = \mathbf{r}_{w} - \mathbf{r}_{i}$, $\rho_{i}$ is computed from Eq.~(\ref{eq:eq2}).

A piston-type wave maker consisting of a group of dummy particles whose displacement is determined by the linear and nonlinear wave maker theory \cite{lee1989transient} can generate regular and irregular waves. Based on the linear wave maker theory, the displacement function of the wave maker $r_{a}$ relies on
\begin{equation}
\left\{
\begin{aligned}
\label{eq:eq19}
{r_{a}} &= E\sin(2\pi f t + \psi) \\
E &= \frac{H(\sinh(2kh) + 2kh)}{\sinh(2kh)\tanh(2kh)}.
\end{aligned}
\right.
\end{equation}
Here, $E$ is the wave stroke, $f$ the wave frequency, $\psi$ the wave phase, $H$ the wave height, and $h$ the water depth. The wave number $k$ followed by the dispersion relation \cite{madsen1971generation}
\begin{equation}
\label{eq:eq20}
\omega^2 = \mathbf{g}k\tanh(kh),
\end{equation}
where $\omega = 2\pi f$ is the wave angular frequency.

For the second-order Stokes wave, the theory by Madsen \cite{madsen1971generation} is adopted, where the wave maker motion is
\begin{equation}
\left\{
\begin{aligned}
\label{eq:eq21}
r_{a} &= -S_{0}\cos(2\pi f t) - S_{0}(\frac{3H\sin(4\pi f t)}{4n_{0}h(4\sinh^{2}(kh) - n_{0} / 2)}) \\
n_{0} &= \frac{1}{2}(1 + \frac{2kh}{\sinh(2kh)}) \\
S_{0} &= \frac{H}{2}\frac{n_{0}}{\tanh(kh)}. 
\end{aligned}
\right.
\end{equation}

The Pierson-Moskowitz and JONSWAP models primarily represent the spectra of irregular waves. The JONSWAP spectrum exhibits a more concentrated wave energy than the Pierson-Moskowitz spectrum, making it more effective in describing the energy distribution of waves in the developmental stage \cite{lopez2019site}. Considering that OWSCs are predominantly placed in nearshore areas, where waves are typically in this developmental stage, it is appropriate to use the JONSWAP spectrum to generate irregular waves \cite{renzi2022application}
\begin{equation}
\left\{
\begin{aligned}
\label{eq:eq22}
S(f) &= \beta_{J}H_{p}^{2}T_{p}^{-4}f^{-5}\exp[-1.25(T_{p}f)^{-4}]\gamma_{J}^{\exp[-\frac{(T_{p}f-1)^2}{(2\delta_{J}^2)}]} \\
\beta_{J} &= \frac{0.0624(1.094 - 0.01915\ln\gamma_{J})}{0.230 + 0.0336\gamma_{J} - 0.185(1.9 + \gamma_{J})^{-1}},
\end{aligned}
\right.
\end{equation}
where $H_{p}$ is the main wave height, $T_{p}$ the peak wave period, $f = \omega/2\pi$ the wave frequency, and $\gamma_{J} = 3.3$ the peak enhancement factor, $\delta_{J}$ dependent on the peak frequency $f_{p} = 1/T_{p}$ \cite{goda2010random}. Our study employed a combination of $N$ random regular waves to simulate nonlinear waves. The wave number $k_{n}$ for each regular wave was consistent with Eq.~(\ref{eq:eq20}). The displacement equation $r_{N}$ can be written as
\begin{equation}
\label{eq:eq23}
r_{N} = \sum_{n=1}^{N}E(f_{n})\cos(2\pi f_{n} t + \psi_{n}).
\end{equation}
Here, $E(f_{n}) = \sqrt{2S_{\omega}(f_{n})\Delta f}$, $f_{1} = 0 \ \text{Hz}$, $f_{N} = 3f_{p}$, $\Delta f = 1/T_{total}$ with $T_{total}$ the total time of simulation, $\psi_{n}$ represents the random phases, and $S_{\omega}(f_{n})$ is defined as
\begin{equation}
\label{eq:eq24}
S_{\omega}(f_{n}) = S(f_{n})(\frac{4\sinh^2(kh)}{2kh + \sinh(2kh)})^{-2}.
\end{equation}

In addition, to prevent numerical divergence caused by excessive movement of the wave maker during the initial computation, we introduced a relaxation time $t_{rex} = 1 \ \text{s}$, and the final wave maker displacement function is as follows
\begin{equation}
r_{N} = 
\left\{
\begin{aligned}
\label{eq:eq25}
&\sin(\frac{\pi t}{2}) r_{N}, & t \leq t_{rex} \\
&r_{N}, & t > t_{rex}.
\end{aligned}
\right.
\end{equation}

Also, to mitigate the impact of wave reflection off the wall on the motion of the OWSC, the wave-particle velocity $\mathbf{v}$ in the damping zone is given by
\begin{equation}
\label{eq:eq26}
\mathbf{v} = \mathbf{v}_{0}(1.0 - \alpha \Delta t (\frac{\mathbf{r} - \mathbf{r}_{0}}{\mathbf{r}_{1} - \mathbf{r}_{0}})).
\end{equation}
$\mathbf{v}_{0}$ the fluid particle velocity at the entrance of the damping zone, the reduction coefficient $\alpha$ is set as 5.0, $\mathbf{r}_{0}$ and $\mathbf{r}_{1}$ are the initial and final position vectors of the damping zone.

\subsection{Numerical validations}
In this section, we primarily verify the accuracy of generating regular and irregular waves and validate the numerical simulations of the 2D and 3D OWSC by comparing them with experimental results.

\subsubsection{Regular waves}
\begin{figure}[htbp]
\centering
\includegraphics[width=\textwidth]{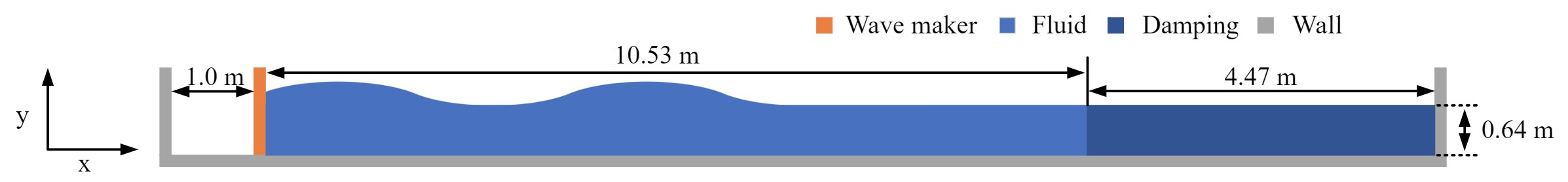}
\caption{The 2D water tank geometry for wave generation verification.}\label{fig2}
\end{figure}

\begin{figure}[htbp]
\centering
\includegraphics[width=\textwidth]{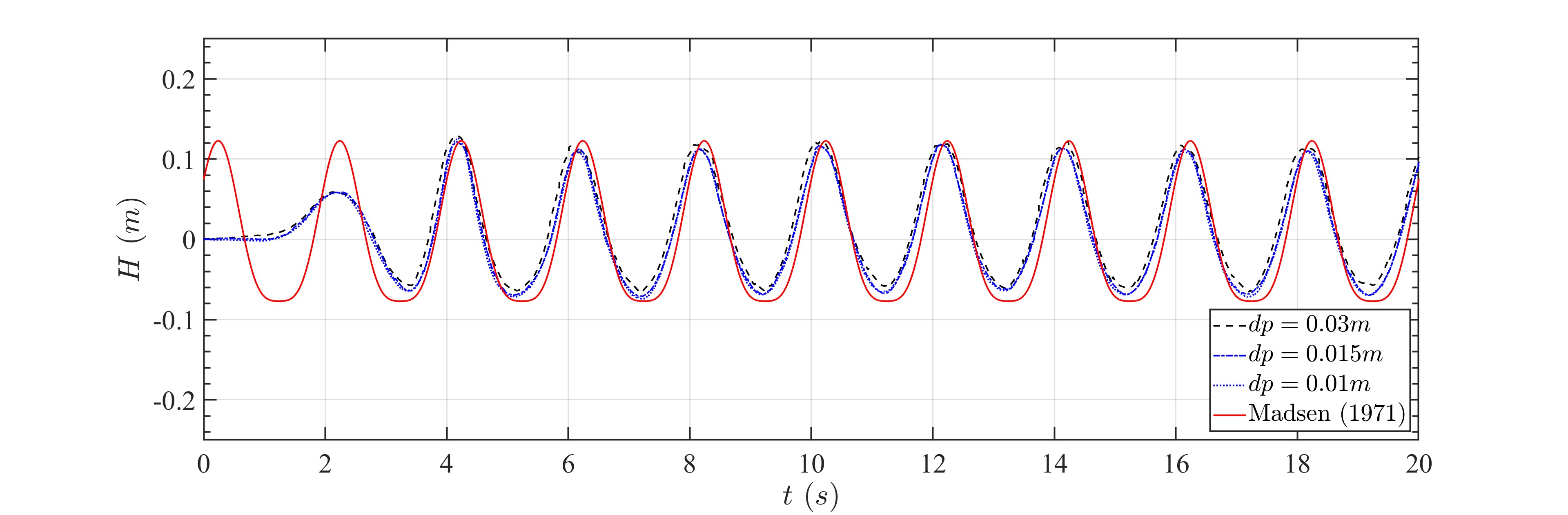}
\caption{Comparison of wave heights under different resolutions at $x = 4.0 \ \text{m}$ with theoretical results.}\label{fig3}
\end{figure}

In this part, we consider the regular wave equation in 2D for validation. The geometry of the numerical simulation is shown in Fig.~\ref{fig2}. The total length of the tank $L$ is 15 m, the water depth $h$ is 0.691 m, and the damping zone is 4.47 m long, the same size as the wavelength $\lambda$. The wave parameters are $H = 0.2 \ \text{m}$ and wave period $T = 2 \ \text{s}$. We place a wave height sensor at a distance of 4 m from the wave maker. As shown in Fig.~\ref{fig3}, our numerical result of wave frequency and amplitude is very close to the analytical solution conducted by Madsen \cite{madsen1971generation}, which indicates that the wave is generated correctly.

Furthermore, we set three different particle resolutions from coarse to fine. The resolution of $dp = 0.015 \ \text{m}$ not only improves the accuracy of the calculation results but also reduces the computation time. Therefore, this resolution will be used for all subsequent 2D simulations.

\subsubsection{Irregular waves}
\begin{figure}[htbp]
    \centering
    \includegraphics[width=\textwidth]{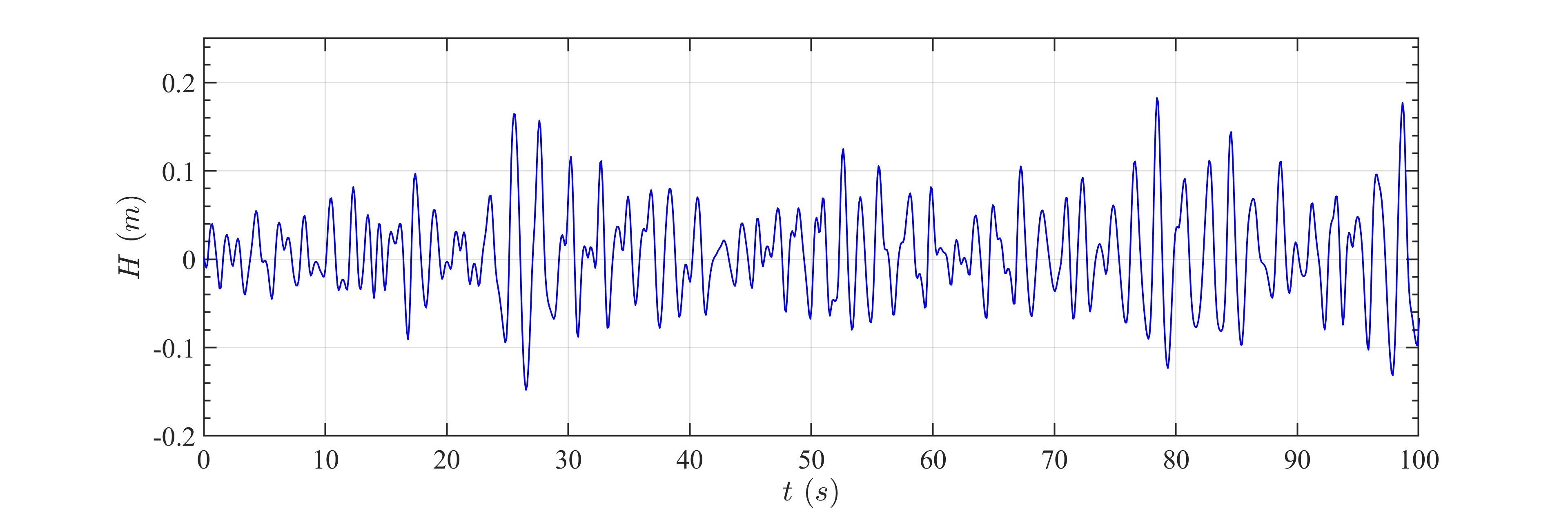}
    \caption{Time series of the free-surface elevation at $x = 0.2 \ \text{m}$ for the irregular wave.}
    \label{fig4}
\end{figure}

\begin{figure}[htbp]
    \centering
    \includegraphics[width=\textwidth]{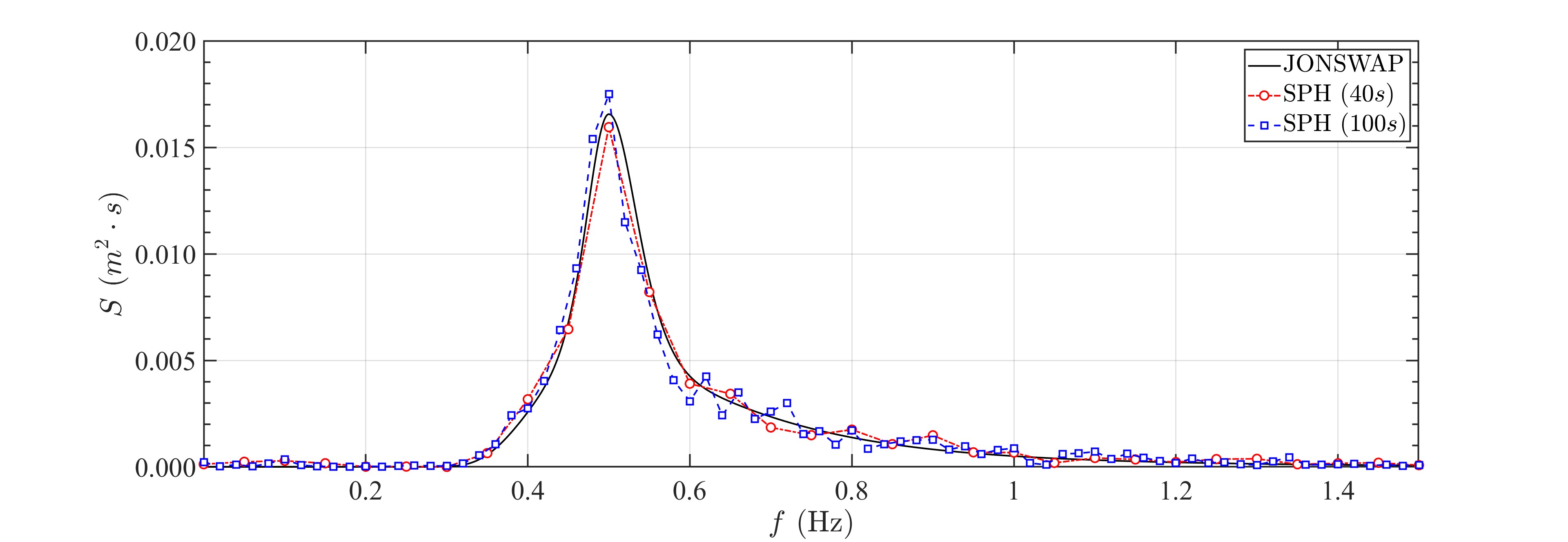}
    \caption{Comparison between the analytical JONSWAP spectrum of Eq.~(\ref{eq:eq22}) and the SPH spectrum corresponding to different time series.}
    \label{fig5}
\end{figure}

For the irregular wace generation, the tank geometry and water depth are the same as regular waves. The typical parameters for irregular waves are $H_{p} = 0.2\ \text{m}$ and $T_{p} = 2.0\ \text{s}$. Note that $T_{total}$ is set for $40\ \text{s}$ and $100\ \text{s}$ for comparison. Fig.~\ref{fig4} presents the wave height time series near the wave maker at $x = 0.2\ \text{m}$. The frequency spectrum is obtained by applying the Fast Fourier Transform (FFT) to the wave height, as shown in Fig.~\ref{fig5}. Notably, our numerical results demonstrate a high degree of agreement with the analytical results derived from the JONSWAP spectrum, and for RL training, $T_{total} = 40\ \text{s}$ is set for one training episode.

\subsubsection{2D and 3D validaitons}
\begin{figure}[htbp]
\centering
\includegraphics[width=\textwidth]{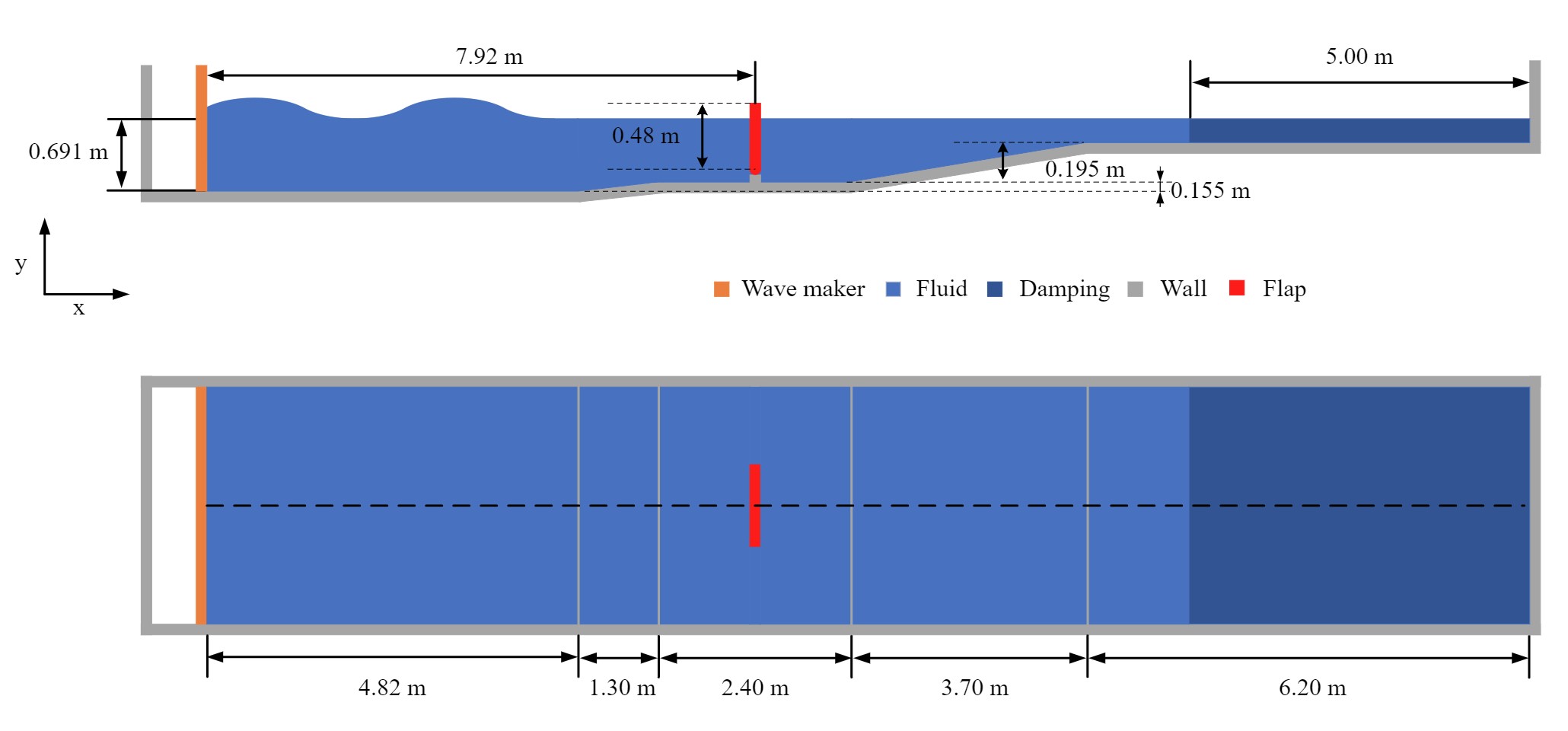}
\caption{Schematic of the wave tank and the OWSC mode.}\label{fig6}
\end{figure}

In this part, the present method for modeling wave interaction with the OWSC is validated in both 2D and 3D. The water tank and OWSC geometries are based on the experiment conducted at the Marine Research Group's hydraulics laboratory at Queen's University Belfast \cite{wei2015wave}, as shown in Fig.~\ref{fig6}. The length, width, and height of the wave tank are $18.4\ \text{m}$, $4.58\ \text{m}$, and $1.0\ \text{m}$. The flap shape of the OWSC is simplified as a box-type with the dimensions of $0.12\ \text{m} \times 1.04\ \text{m} \times 0.48\ \text{m}$. It is set at $7.92\ \text{m}$ far from the wave maker in the $x$ direction and the middle in the $z$ direction. The water depth $h$ is $0.691\ \text{m}$, and the hinge height is $0.16\ \text{m}$. The mass and inertia of the flap are $33\ \text{kg}$ and $1.84\ \text{kg} \cdot \text{m}^2$. The damping coefficient $k_{d}$ is set to 20. In addition, we set three wave height sensors in the $x$ direction, which are WP04 (3.99 m), WP05 (7.02 m), and WP12 (8.82 m).

\begin{figure}[htbp]
    \centering
    \begin{subfigure}{\textwidth}
        \centering
        \includegraphics[width=\textwidth]{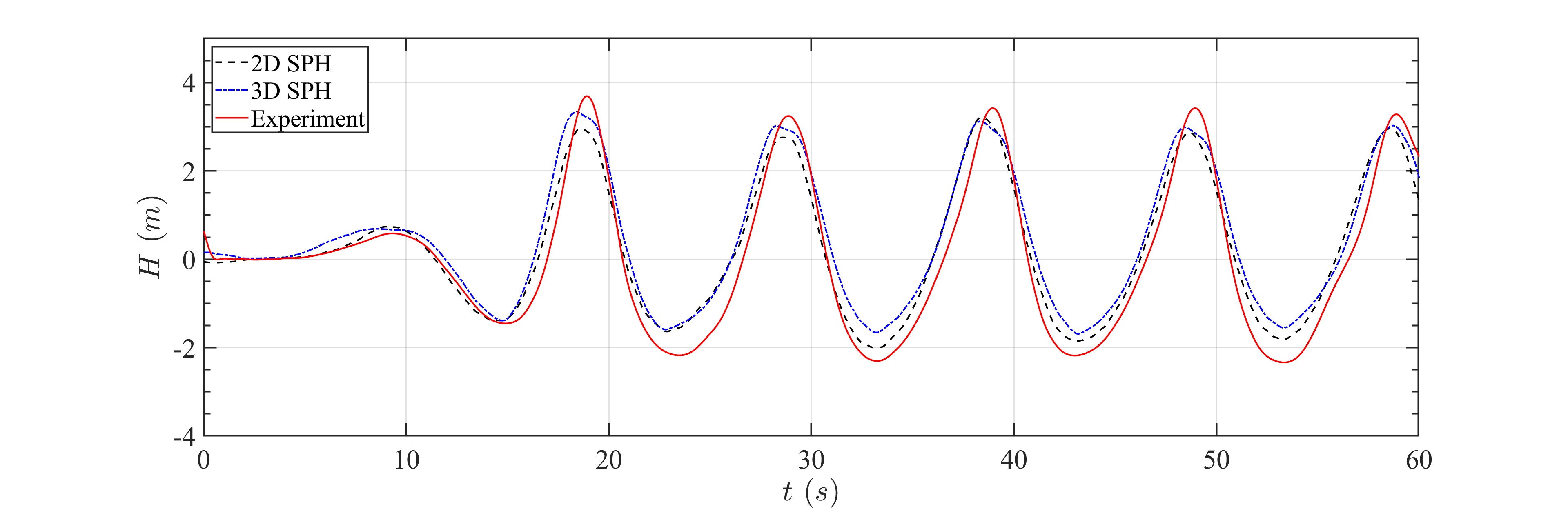}
        \caption{WP04}
        \label{fig7a}
    \end{subfigure}
    
    \begin{subfigure}{\textwidth}
        \centering
        \includegraphics[width=\textwidth]{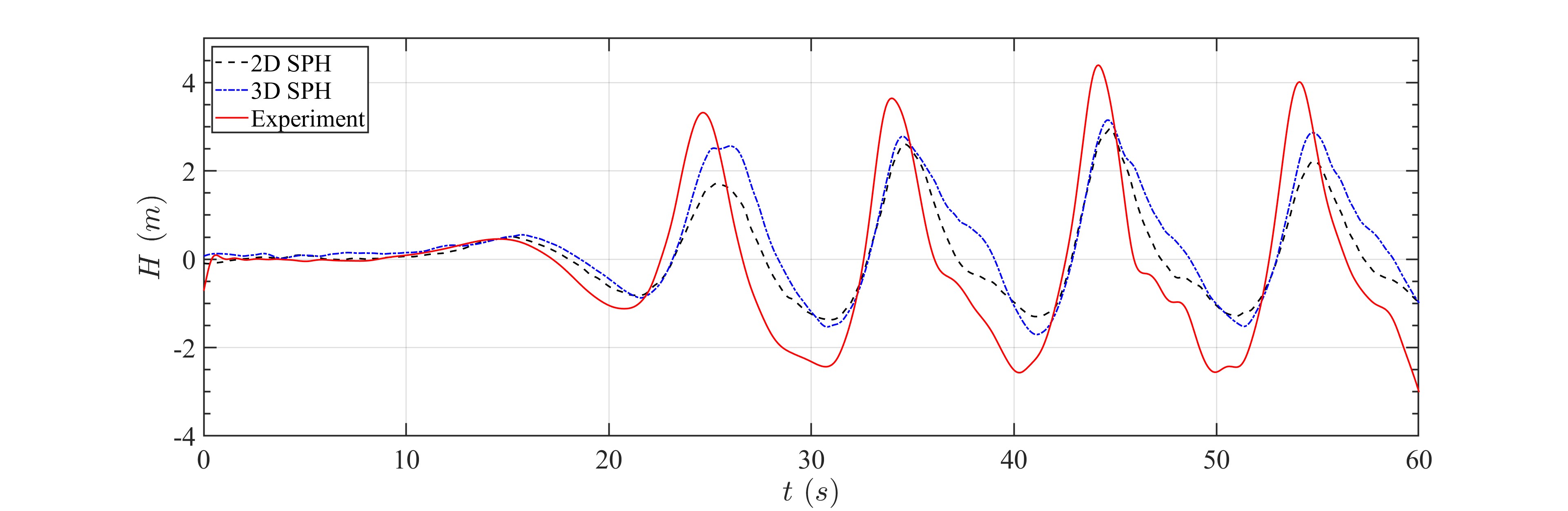}
        \caption{WP05}
        \label{fig7b}
    \end{subfigure}
    
    \begin{subfigure}{\textwidth}
        \centering
        \includegraphics[width=\textwidth]{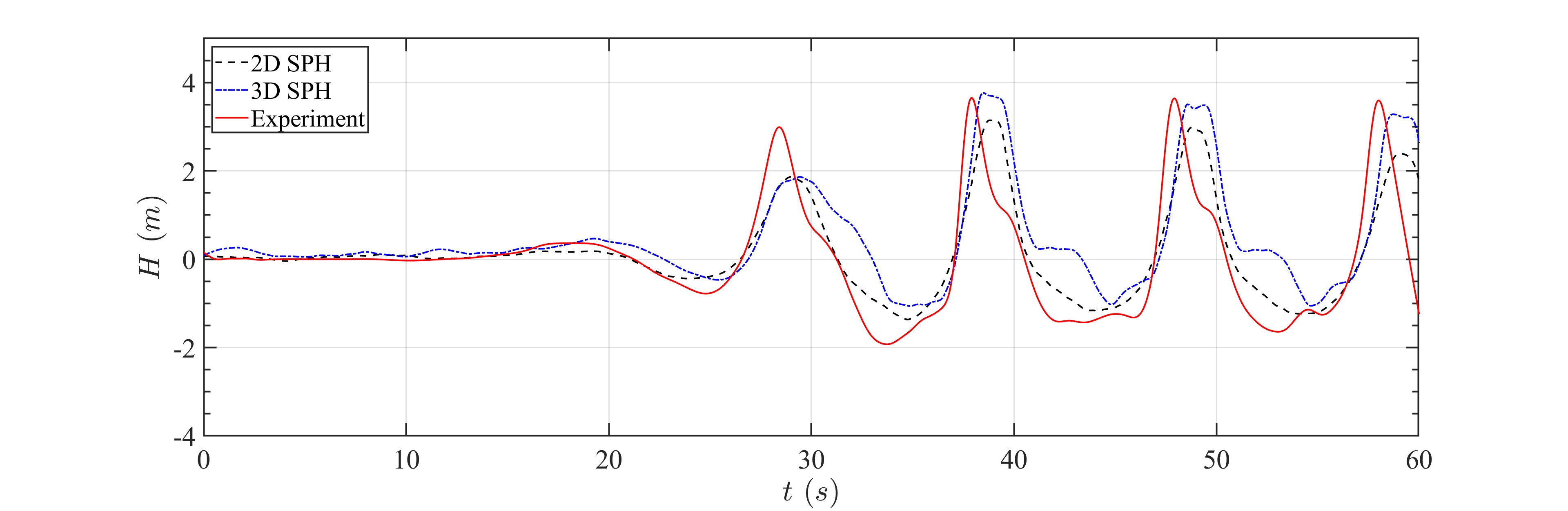}
        \caption{WP12}
        \label{fig7c}
    \end{subfigure}
    
    \caption{Comparison of free surface elevations for wave height $H=5.0 \ \text{m}$ and wave period $T = 10\ \text{s}$ at different wave probes. (a) $x = 3.99\ \text{m}$ for WP04, (b) $x = 7.02\ \text{m}$ for WP05, and (c) $x = 8.82\ \text{m}$ for WP12}
    \label{fig7}
\end{figure}

\begin{figure}[htbp]
\centering
\includegraphics[width=\textwidth]{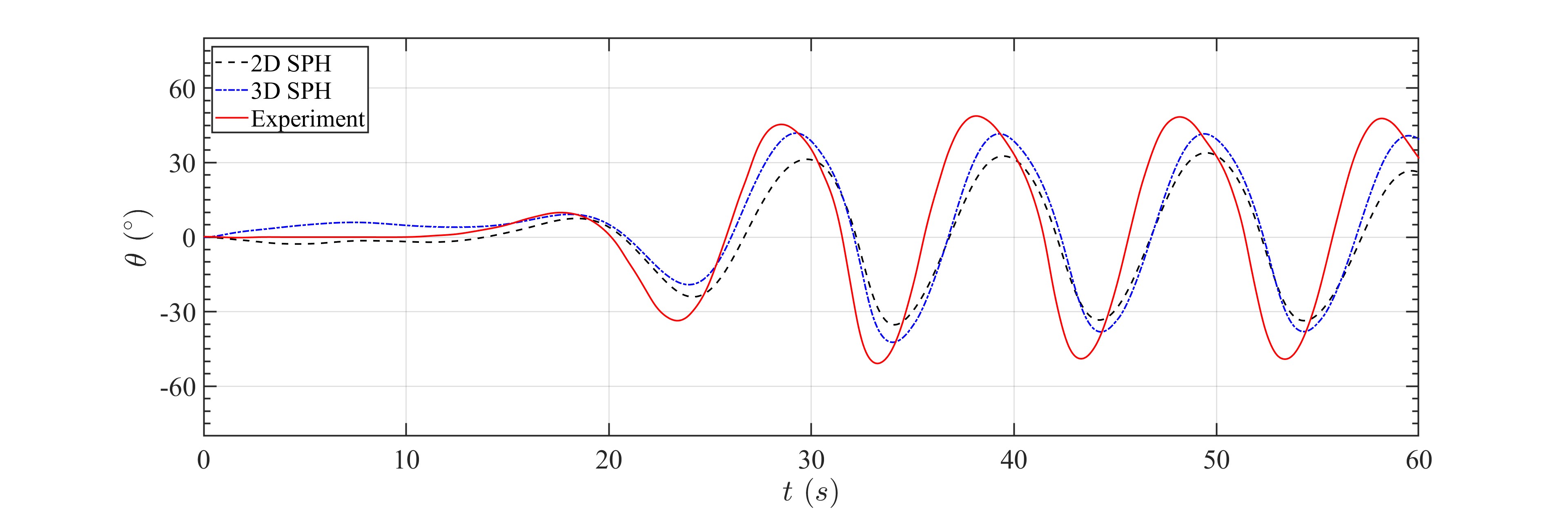}
\caption{Comparison of the flap rotation.}\label{fig8}
\end{figure}

In the current simulation, the wave maker creates a simple harmonic wave with the wave height $H=0.2\ \text{m}$ and wave period $T = 2\ \text{s}$. The simulation parameter is $ 1:25 $ in scale, and both 2D and 3D results presented herein have been converted to full scale as in Ref. \cite{zhang2021efficient}. The 2D simulation is computed on a Mac OS system, with an Apple M1 Max core and 32GB RAM, while the 3D simulation is carried on a Windows system, with an AMD Ryzen 9 7950X and 48GB RAM. The total fluid particle numbers are 40027 and 1.542 million, for 2D and 3D simulations, respectively. From Fig.~\ref{fig7}, we can see that WP04 is far away from the flap, so the interaction of the wave and flap has little influence on the wave height, and our results show that the wave maker can generate an accurate sine wave with minor errors. WP05 and WP12 are set near the flap, and we can find that our simulations can capture the influence of interaction on wave height in both places. Furthermore, by comparing the wave heights at WP05 and WP12, it is evident that the wave height significantly decreases after passing through the OWSC, indicating an energy reduction. This reduction indirectly demonstrates that a portion of the energy has been converted into wave energy. Fig.~\ref{fig8} shows that the flap rotation simulation results are consistent with the experiment. The 3D results are much better, where antisymmetric diffracted waves traveling in all directions, including tangentially to the flap, can induce near-resonant phenomena that enhance the exciting torque on the converter \cite{renzi2014wave}.

The results demonstrate that our simulations based on SPHinXsys are suitable for setting the physical environment for RL. Notably, a 2D simulation requires only 3.5 minutes of computing time for 12 seconds of physical time calculation, whereas a 3D simulation demands 3.5 hours. Given that each DRL training session necessitates at least 200 episodes, we employ 2D simulations for training and utilize 3D simulations for policy validation.

\section{Direct deep reinforcement learning}
\begin{figure}[htbp]
\centering
\includegraphics[width=\textwidth]{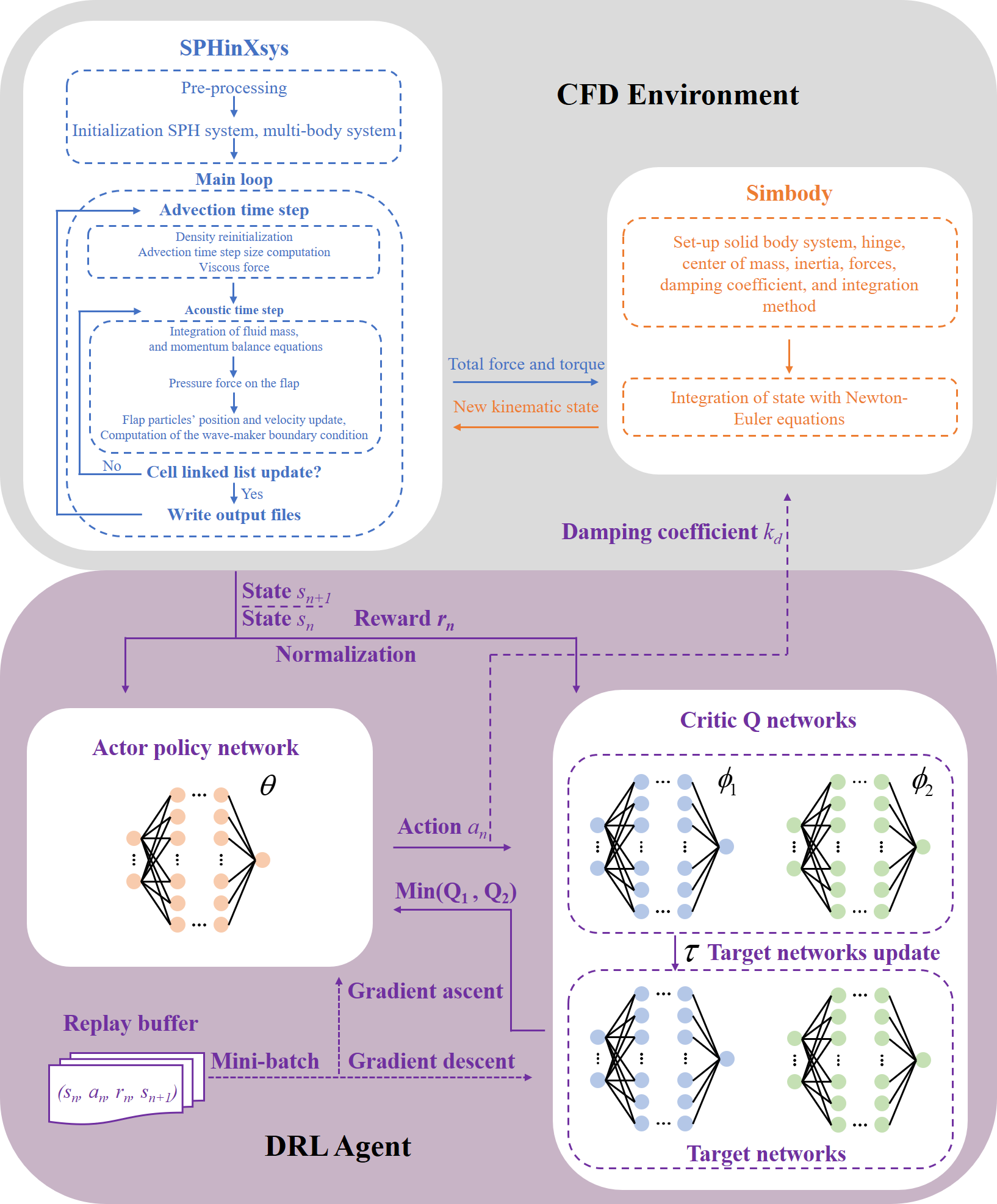}
\caption{The current CFD-DRL framework encompasses an integrated interaction process between two primary components: the CFD Environment and the DRL Agent.}\label{fig9}
\end{figure}

Fig.~\ref{fig9} provides an overview of the CFD-DRL platform. The DRL training process consists of two key components: the environment and the agent. The previous chapter details the CFD environment. Here, the focus shifts to the agent. The sampled state vector from SPHinXsys is normalized and passed to the agent. The reward is calculated based on the change in state between two consecutive time steps and the variation in reward parameters during the action application. All experiences are uniformly stored in the replay buffer for future use. We show a typical framework of the DRL algorithm: soft actor critics (SAC), which takes actor-critic architecture \cite{haarnoja2018soft}. The policy network (actor) outputs actions fed into Simbody and critic networks in real-time. The critic networks evaluate the quality of these actions, and the smaller $Q_{i}$ is chosen to update the policy network with gradient ascent. The critic networks will be updated with target networks using gradient descent.

\subsection{DRL environments}
The numerical simulation environment, corresponding observation probes, and action transition functions are created using SPHinXsys. Pybind11 is utilized to package the entire computational module as dynamic link libraries (DLLs) or shared objects (SOs), which can then be imported and executed directly within the Python environment \cite{pybind11}.

A standard DRL environment is built based on OpenAI's Gym library \cite{brockman2016openai}, which includes two essential functions: reset and step. In the reset function, the SPHinXsys-based numerical simulation environment is initialized, and the initial observations are collected using probes. The step function receives action values from the DRL algorithm, passes them to SPHinXsys for numerical calculations under the current action, collects new observations, and calculates the reward.

\begin{figure}[htbp]
\centering
\includegraphics[width=0.6\textwidth]{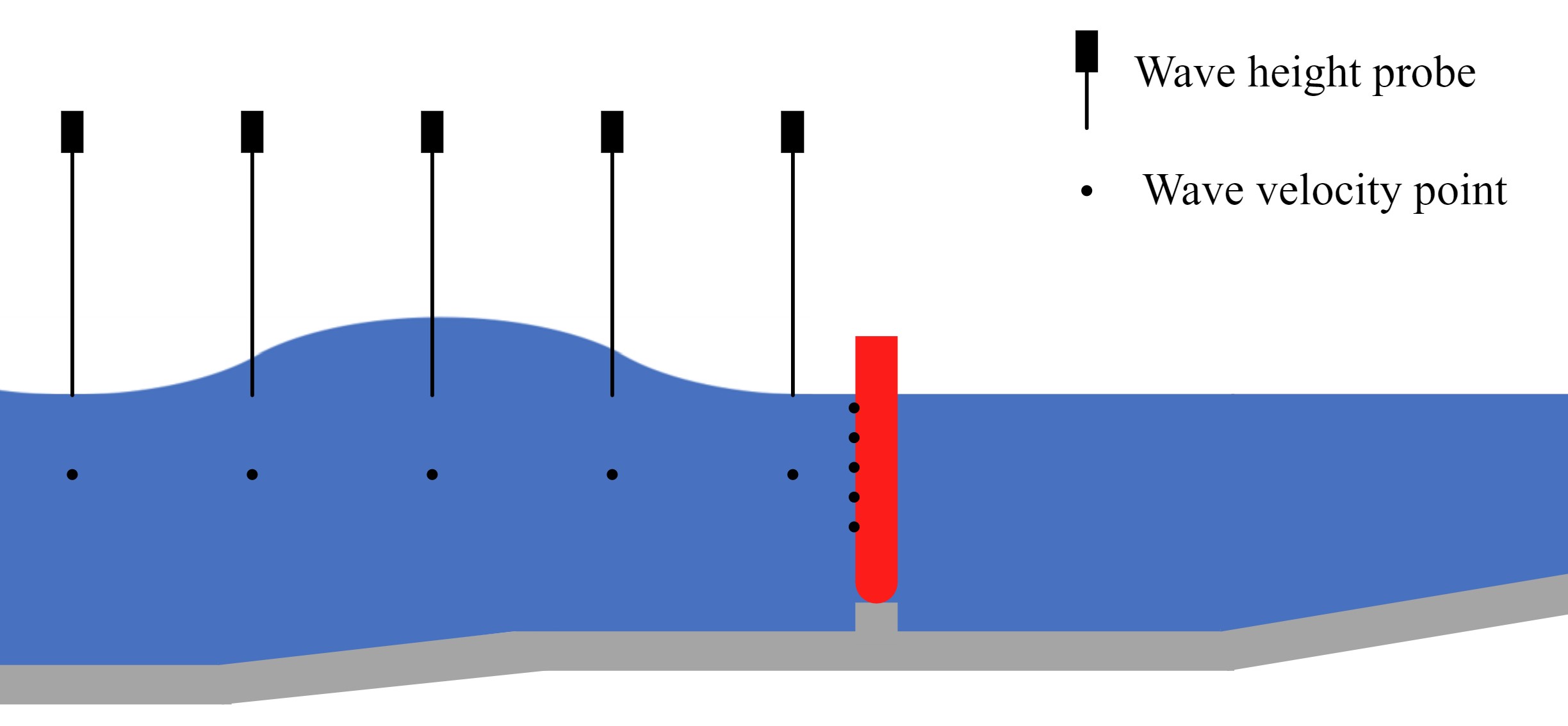}
\caption{Two main components of the observation vector.}\label{fig10}
\end{figure}

Rabault et al. \cite{rabault2019artificial} shown that more observations will give the agent more information to update its network and improve the training result. Therefore, 38 observations are set as the state vector $s_{n}$ from the environment to capture the structure of the regular wave and its impact on the flap. The components of the observations are based on two parts, as shown in Fig.~\ref{fig10}. The first part is the wave properties, including velocity and wave height at five positions, starting from $x = 3.0\ \text{m}$ to $x = 7.6\ \text{m}$, which is close to one wavelength. Another 5 points from $y = 0.3\ \text{m}$ to $y = 0.7\ \text{m}$ on the front panel of the OWSC device are also set to get the wave velocity. The second part is the characteristics of the flap, including flap rotation and angular velocity, and damping coefficient $k_{d}$ of the PTO system. The x-axis and y-axis total force on the flap are not considered as they fluctuate broadly with too much noise. Besides, it is ideal that the observations from 2D and 3D simulations can be close. Thus, the observation positions are set on the middle plane in the z-axis as shown in Fig.~\ref{fig6}, and only x-axis and y-axis properties are considered.

In the present work, the action $a_{n}$ in one action time step $t_{a} = 0.1\ \text{s}$ of variation in the damping coefficient $|\Delta k_{d}| \leq 25$. However, directly changing the damping coefficient has the potential risk to the computation divergence.Therefore, the current numerical damping coefficient $k_{d}^{n}$ and the subsequent numerical damping coefficient $k_{d}^{n+1}$ have a simple linearly increasing change of $\Delta k_{d}/M$ during the time step of $t_{a}/M$. Considering that the $\Delta t_{ac} = 0.00021 \ \text{s}$, $M = 10$ is sufficient to ensure accuracy in the 2D simulation. Note that a large damping coefficient will make the flap's rotation challenging. A small damping coefficient will result in larger deflection angles, which may cause the OWSC break in real applications. In present work, $0 \leq k_{d} (N \cdot m \cdot s / rad) \leq 100$ is applied. As observed in Fig.~\ref{fig8}, we can see that even if the damping coefficient is set to 0, the most extensive rotation of the flap is 50\textdegree, which is smaller than the critical value. Although the entire OWSC can still run under extreme boundary conditions, a penalty term is given in the reward as $P^{*} = -1 \cdot \left[ (k_d + \Delta k_d < 0) \lor (k_d + \Delta k_d > 100) \right]$. 

Based on Senol and Raessi's work \cite{senol2019enhancing}, the instance energy harvesting in one action time step can be defined as
\begin{equation}
\label{eq:eq27}
P_{take-off} = \sum_{n=0}^{M - 1}k_{d}^{n}(\frac{\Omega_{n+1} + \Omega_{n}}{2})^{2}. 
\end{equation}
Here, $\Omega_{n}$ is the flap angular velocity. It is evident that larger $P_{take-off}$ means more instantaneous wave energy harvesting by the OWSC. The study of Zhang et al. \cite{zhang2021efficient} shows that the average wave energy harvesting factor (CF) of 3D OWSCs can reach a peak when $k_{d}^{*}$ is around 40 N $\cdot$ m $\cdot$ s / rad. Combined with Table~\ref{table1}, the verification results show that in the case of 2D OWSC, the total wave energy conversion reaches a maximum value at $k_{d}^{*} = 45 N \cdot m \cdot s / rad$. Then, the instance energy harvesting in one action time step with $k_{d}^{*}$ is recorded as the baseline $P_{baseline}$ in the reward to help the agent distinguish good from harmful actions. The final reward can be calculated as
\begin{equation}
\label{eq:eq28}
r_{n} = P_{take-off} - P_{baseline} + P^{*}.
\end{equation}

\begin{table}[htbp]
    \centering
    \caption{The variations of the total wave energy conversion $E_t$ in terms of damping coefficients.}
    \label{table1}
    \resizebox{\textwidth}{!}{
    \begin{tabular}{cccccccccccc}
        \toprule
        \(k_d\) (N $\cdot$ m $\cdot$ s / rad) & 10 & 20 & 30 & 40 & 45 & 50 & 60 & 70 & 80 & 90 & 100 \\ \midrule
        \(E_{t}\) (J) & 195.19 & 281.38 & 309.56 & 321.26 & 321.96 & 316.52 & 310.86 & 300.92 & 291.02 & 279.37 & 267.47 \\ \bottomrule
    \end{tabular}
    }
\end{table}

\subsection{DRL algorithms}
RL algorithms can be broadly categorized into two types: on-policy and off-policy. On-policy algorithm updates its policy network after one episode and uses the newly updated policy to collect data in the next episode. The typical algorithm is proximal oolicy optimization (PPO) \cite{schulman2017proximal}. On the other hand, the policy for updating and collecting is different for the off-policy algorithm. The typical methods are twin delayed deep deterministic policy gradient (TD3) \cite{fujimoto2018addressing} and SAC. 

More specifically, the action value function $Q_{\pi_\theta}(s_{n}, a_{n})$ and the state value function $V_{\pi_\theta}(s_{n})$ can be defined as: 
\begin{equation}
\left\{
\begin{aligned}
\label{eq:eq29}
Q_{\pi_\theta}(s_{n}, a_{n}) &= \mathbf{E}_{\pi_\theta}[\sum_{t=n}^{\infty} (\gamma_{t} r_{t} | s_{t}, a_{t})] \\
V_{\pi_\theta}(s_{n}) &= \mathbf{E}_{a_{n} \sim \pi_\theta}[\sum_{t=n}^{\infty} (\gamma_{t} r_{t} | s_{t})],
\end{aligned}
\right.
\end{equation}
where $\pi_{\theta}$ is the policy network with parameter $\theta$, $\gamma_{t} \in [0, 1]$ the discount factor for the future reward, and $\sum_{t=n}^{\infty} \gamma_{t} r_{t}$ usually defined as return $G_{n}$. The policy network plays an important role in the optimization process, as its input is the current state $s_{n}$, and its output is the action $a_{n}$ or the probability density function of the action $\pi_{\theta}(\cdot | s_{n})$. The action value function $Q_{\pi_\theta}(s_{n}, a_{n})$ represents the expected return obtained by taking action $a_{n}$ in the current state $s_{n}$ and then following the policy $\pi_\theta$. The state value function $V_{\pi_\theta}(s_{n})$ represents the expected return obtained by following the policy $\pi_\theta$ and the current state $s_{n}$. 

\subsubsection{The PPO algorithm}
The core of the PPO algorithm is to build an objective function $J$ to characterize the return $G_{n}$ under the parameter $\theta$ of the current policy network. The best policy and return are obtained by updating $\theta$ to maximize the objective function. Based on the Policy Gradient Theorem (PGT), the objective function $J(\theta)$ can be written as \cite{schulman2017proximal}
\begin{equation}
\label{eq:eq30}
J(\theta) = \mathbf{E}_{s_{n} \sim \mathcal{D}}[\mathbf{E}_{a_{n} \sim \pi_{\theta_k}}[\frac{\pi_{\theta}(a_{n}|s_{n})}{\pi_{\theta_{k}}(a_{n}|s_{n})}\cdot A^{\pi_{\theta_k}}(s_{n}, a_{n})]],
\end{equation}
where $\mathcal{D}$ is the replay buffer, $\theta_{k}$ the old parameter of the policy network. Also, $A^{\pi_{\theta_k}}(s_{n}, a_{n})$ is the advantage function which can be defined as
\begin{equation}
\label{eq:eq31}
A^{\pi_{\theta_k}}(s_{n}, a_{n}) = r_{n} + \gamma V^{\pi_{\theta_k}}_{\phi}(s_{n + 1}) - V^{\pi_{\theta_k}}_{\phi}(s_{n}),
\end{equation}
where $V^{\pi_{\theta_k}}_{\phi}(s_n)$ is the estimated value of state $s_n$ by the critic network with parameters $\phi$, the superscript $\pi_{\theta_k}$ indicates that the data is collected using the policy employed during the $k$-th iteration, and $\gamma$ is the discount factor. The advantage function will not affect the expectation but improve the policy's performance \cite{schulman2015high}. A key feature of PPO is the clipped surrogate objective, which is designed to prevent huge policy updates. The objective function of $J(\theta)$ is then rewritten as follows
\begin{align}
\label{eq:eq32}
J(\theta) &= \mathbf{E}_{s_{n} \sim \mathcal{D}} \left[ \mathbf{E}_{a_{n} \sim \pi_{\theta_k}} \left[ \min \left( r_{\theta}(s_{n}, a_{n}), \right. \right. \right. \notag \\
& \left. \left. \left. \text{clip}(r_{\theta}(s_{n}, a_{n}), 1 - \sigma, 1 + \sigma) \right) \cdot A^{\pi_{\theta_k}}(s_{n}, a_{n}) \right] \right],
\end{align}
where $r_{\theta}(s_{n}, a_{n}) = \pi_{\theta}(a_{n}|s_{n}) / \pi_{\theta_{k}}(a_{n}|s_{n})$, and $\sigma = 0.2$. Note that the objective function $J(\theta)$ is only relevant with the policy network parameter $\theta$, and our target is to update the policy by maximizing $J(\theta)$ with stochastic gradient ascent with Adam scheme\cite{kingma2014adam}:
\begin{equation}
\label{eq:eq33}
\theta_{k + 1} \leftarrow \theta_{k} + \alpha \nabla_\theta J(\theta),
\end{equation}
where $\alpha$ is the learning rate.

The critic network $V_{\phi}$ in PPO is primarily used to represent the state value function and its loss function is defined as
\begin{equation}
\left\{
\begin{aligned}
\label{eq:eq34}
L(\phi) &= \mathbf{E}_{s_{n} \sim \mathcal{D}} \left[ \left( V_{\phi}(s_{n}) - R_n \right)^2 \right] \\
R_n &= r_n + \gamma V_{\phi_{k}}(s_{n+1}).
\end{aligned}
\right.
\end{equation}
To update the critic network parameters $\phi$, the value function loss is minimized using gradient descent as
\begin{equation}
\label{eq:eq35}
\phi_{k+1} \leftarrow \phi_{k} - \alpha_0 \nabla_\phi L(\phi).
\end{equation}

\subsubsection{The TD3 algorithm}
The TD3 algorithm is based on policy gradient methods, where the policy network $\pi_{\theta}$ outputs actions $a_{n}$ directly. The value networks $Q_{\phi_i}$, called critics, provide value estimates based on $s_{n}$ and $a_{n}$ suggested by the policy network. TD3 employs two value networks to mitigate overestimation issues and obtains more reliable value estimates using the minimum value predicted by the two networks \cite{fujimoto2018addressing}. Each network (the policy and the two value networks) is paired with a corresponding target network. Therefore, a total of six networks are utilized during the training process. 

The objective function of a TD3 policy network is defined as
\begin{equation}
\label{eq:eq36}
J(\theta) = \mathbf{E}_{s_{n} \sim \mathcal{D}} \left[ Q_{\phi_1}(s_{n}, \pi_{\theta}(s_{n})) \right].
\end{equation}

Also the loss for the critic networks is given by temporal difference (TD)
\begin{equation}
\label{eq:eq37}
L(\phi_i) = \mathbf{E}_{s_{n} \sim \mathcal{D}} \left[ \left( Q_{\phi_i}(s_{n}, a_{n}) - y_{n} \right)^2 \right], i = 1, 2.
\end{equation}
Here, the target value $y_n$ is calculated based on the Bellman Equation (BE)
\begin{equation}
\label{eq:eq38}
y_{n} = r_{n} + \gamma \min_{i=1,2} Q_{\phi_i'}(s_{n+1}, \pi_{\theta'}(s_{n+1}) + \epsilon),
\end{equation}
where $Q_{\phi_i'}$ means the target critic network, $\pi_{\theta'}$ is the target policy network and $\epsilon$ is the truncated Gaussian noise.

Gradient ascent on the policy network and gradient descent on the critic networks are performed to update their corresponding parameters. Note that the update frequency of the policy network and three target networks is generally lower than the critic networks.

\subsubsection{The SAC algorithm}
SAC consists of a policy network that outputs an action probability density function, two value networks, and two target networks corresponding to the value networks. 

In the PPO state space exploration primarily relies on sampling from the action probability distribution output by the policy network, whereas TD3 achieves exploration by artificially adding noise to the output of the action. Compared with the PPO and TD3 algorithms, the SAC algorithm incorporates the entropy of the policy into the state-value function, encouraging exploration by maximizing the return regularized by entropy
\begin{equation}
\label{eq:eq39}
V_{\pi_\theta}(s_{n}) = \mathbf{E}_{a_{n} \sim \pi_\theta}[\sum_{t=n}^{\infty} (\gamma_{t} r_{t} | s_{t}) + \beta\mathcal{H}(\pi_{\theta}(s_{n}))],
\end{equation}
where $\beta$ is the regularization coefficient.

The objective function of the policy network is
\begin{equation}
\label{eq:eq40}
J(\theta) = \mathbf{E}_{s_n \sim \mathcal{D}} \left[\min\limits_{i=1,2} Q_{\phi_{i}}(s_n, \tilde{a}_{n}) - \beta \log \pi_\theta(\tilde{a}_{n} | s_n) \right],
\end{equation}
where $\tilde{a}_{n}$ is the sample from $\pi_{\theta}(\cdot | s_{n})$.

Also, the loss for the critic networks is also calculated by TD, while with a different definition of $y_{n}$
\begin{equation}
\left\{
\begin{aligned}
\label{eq:eq41}
L(\phi_i) &= \mathbf{E}_{s_{n} \sim \mathcal{D}} \left[ \left( Q_{\phi_i}(s_{n}, a_{n}) - y_{n} \right)^2 \right], i = 1, 2 \\
y_{n} &= r_{n} + \gamma \left( \min_{i=1,2} Q_{\phi_i'}(s_{n+1}, \tilde{a}_{n+1}) - \beta \log \pi_{\theta}(\tilde{a}_{n+1} | s_{n+1}) \right),
\end{aligned}
\right.
\end{equation}
where $\tilde{a}_{n+1}$ is the sample from $\pi_{\theta}(\cdot | s_{n+1})$.

\section{Result}
In this section, various RL algorithms for optimizing the energy harvesting of the PTO system of an OWSC under regular wave conditions will be first analyzed. Then, the generalization capability of 2D training policy in 3D simulations will be verified. Subsequently, the training of agents under irregular wave conditions is investigated. Last but not least, optimizing the damping coefficient in a strongly nonlinear system comprising a dual OWSC system is explored.

\subsection{Study of DRL algorithms}
Initially, three typical DRL algorithms are trained:i.e., PPO, TD3, and SAC. The parameters of the policy and critic networks under the three algorithms are consistent, with two hidden layers and 512 neurons in each layer. Other settings of the neural network and algorithm hyperparameters are shown in Table~\ref{table2}.
\begin{table}[htbp]
    \centering
    \caption{Basic hyperparameters of different DRL algorithms.}
    \label{table2}
    \begin{tabular}{@{}lccc@{}}
        \toprule
        Algorithm & PPO & TD3 & SAC \\ \midrule
        Activation function & \(\tanh\) & \(\tanh\) & \(\tanh\) \\ 
        Learning rate (\(\alpha\)) & 3e-4 & 3e-4 & 1e-3 \\ 
        Steps per epoch & 2048 & 2048 & 2048 \\ 
        Batch size & 256 & 256 & 256 \\ 
        Discount factor (\(\gamma\)) & 0.99 & 0.99 & 0.99 \\ 
        Soft update (\(\tau\)) & - & 0.005 & 0.005 \\ \bottomrule
    \end{tabular}
\end{table}

\begin{figure}[htbp]
    \centering
    \includegraphics[width=\textwidth]{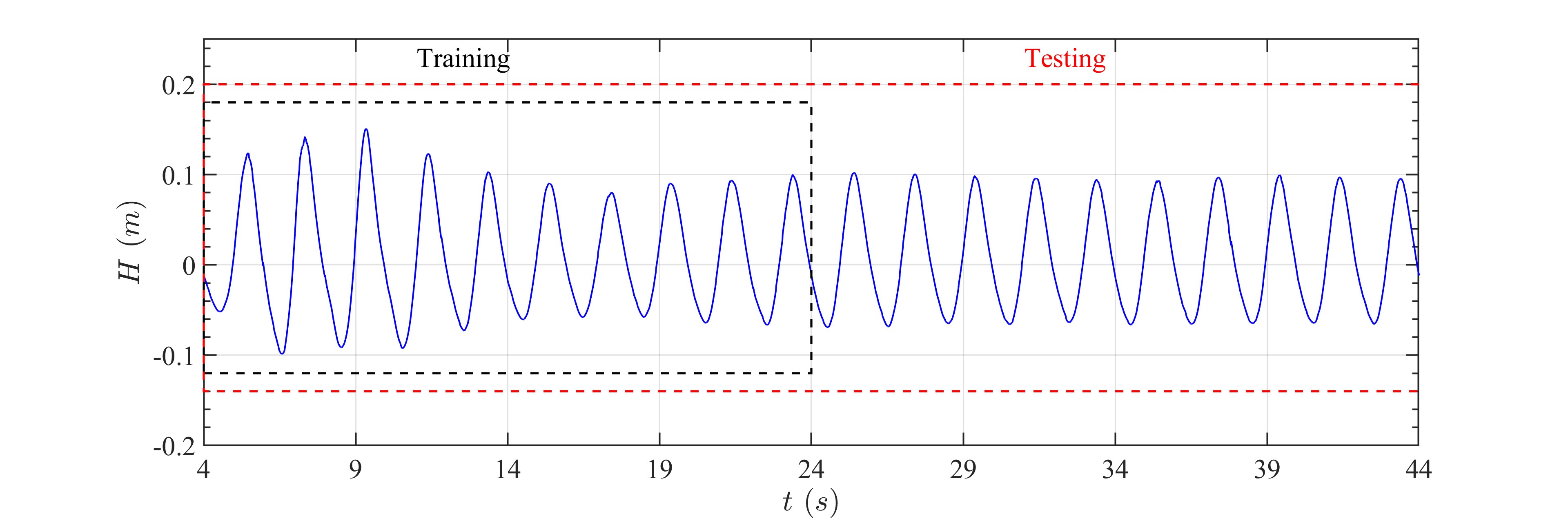}
    \caption{Wave height in front of the flap.}
    \label{fig11}
\end{figure}

Fig.~\ref{fig11} illustrates the wave height at $x = 7.6 \ \text{m}$ in front of the flap. It is observed that around the 4-second mark, the wave reaches and interacts with the flap. Consequently, the training phase is initiated at the 4-second mark and continues until the 24-second mark, encompassing 20 seconds and 200 actions. The testing phase extends over 40 seconds, incorporating 400 actions. 
\begin{figure}[htbp]
    \centering
    \includegraphics[width=0.5\textwidth]{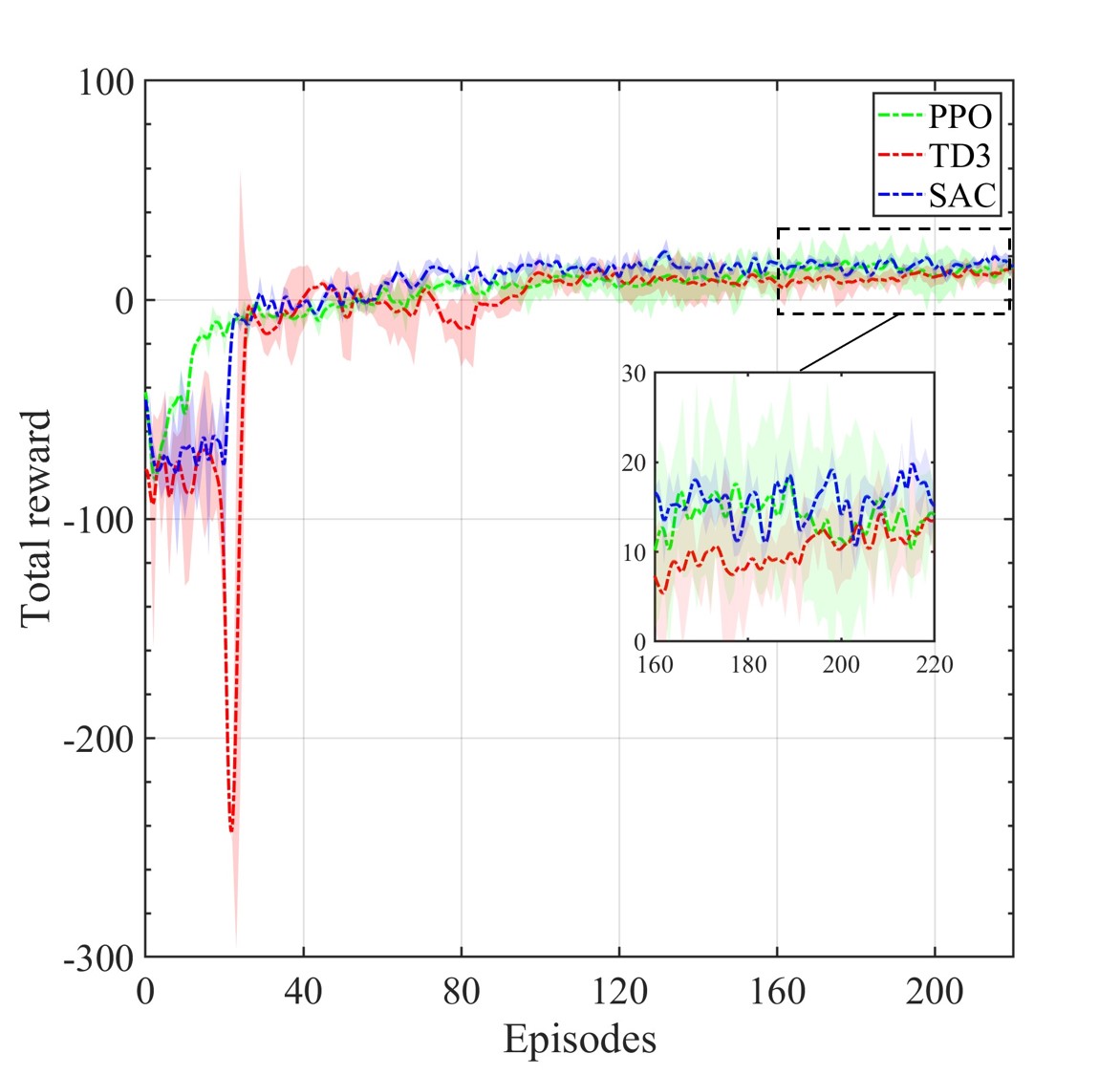}
    \caption{Total reward curves in the training process with three agents.}
    \label{fig12}
\end{figure}

The overall reward curves with standard deviation shadows are depicted in Fig.~\ref{fig12}. The training consisted of 220 episodes, with TD3 and SAC requiring the pre-collection of data for the first 20 episodes, a process of gathering initial data before the actual training. Pre-collection resulted in oscillations around -80 in the reward curves during this phase. However, both algorithms quickly identified effective energy enhancement strategies within the first ten episodes after training commenced. Considering TD3 incorporates noise artificially to enrich exploration, it displayed instability and only began to stabilize after 90 episodes.

\begin{figure}[htbp]
    \centering
    \includegraphics[width=\textwidth]{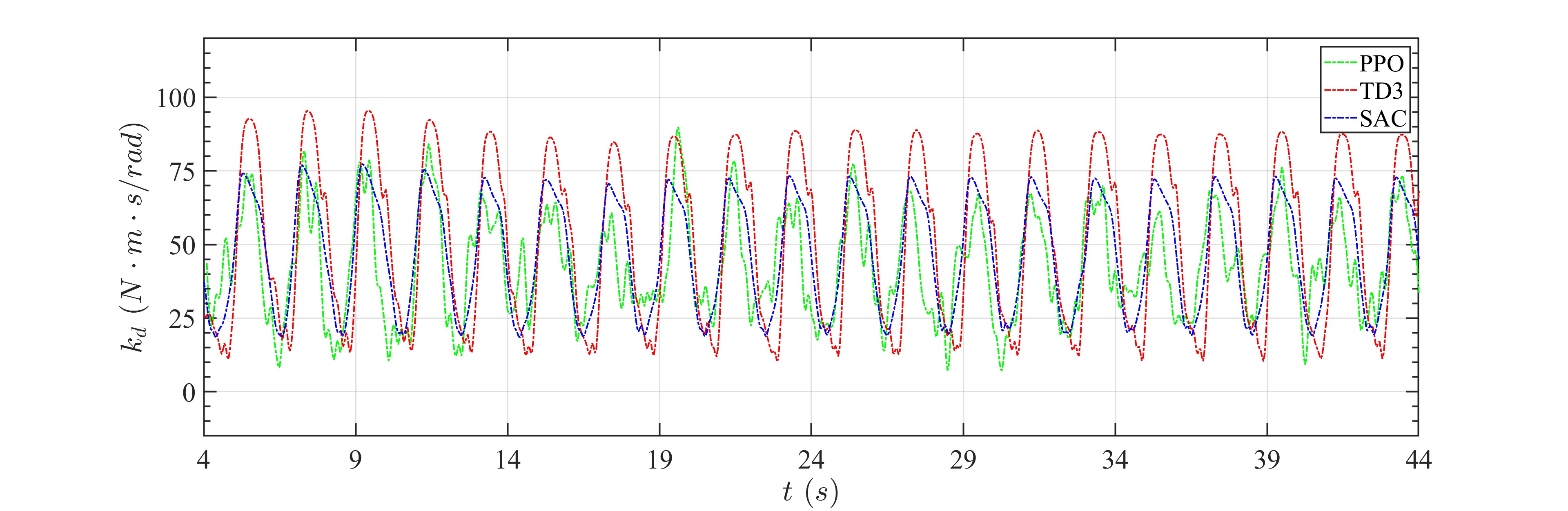}
    \caption{Adaptive damping coefficients of the PTO system with different agents.}
    \label{fig13}
\end{figure}

On the other hand, SAC explored and converged to optimal strategies by approximately 60 episodes with entropy regularization. PPO, not requiring initial data collection, showed a slower but steady improvement from the beginning, converging to effective strategies around 80 episodes. From the reward trends observed between episodes 160 and 220, SAC demonstrates its best performance with the slightest standard deviation, indicating superior stability over other two algorithms. In addition, since the incident wave is regular, the dynamic curve of the damping coefficient is also periodic, as shown in Fig.~\ref{fig13}. It can be observed that the overall fluctuation of the PPO algorithm is substantial, whereas the TD3 algorithm converges to a locally optimal solution due to insufficient exploration. Therefore, SAC will be applied for subsequent agent training.

\begin{figure}[htbp]
\centering
\includegraphics[width=\textwidth]{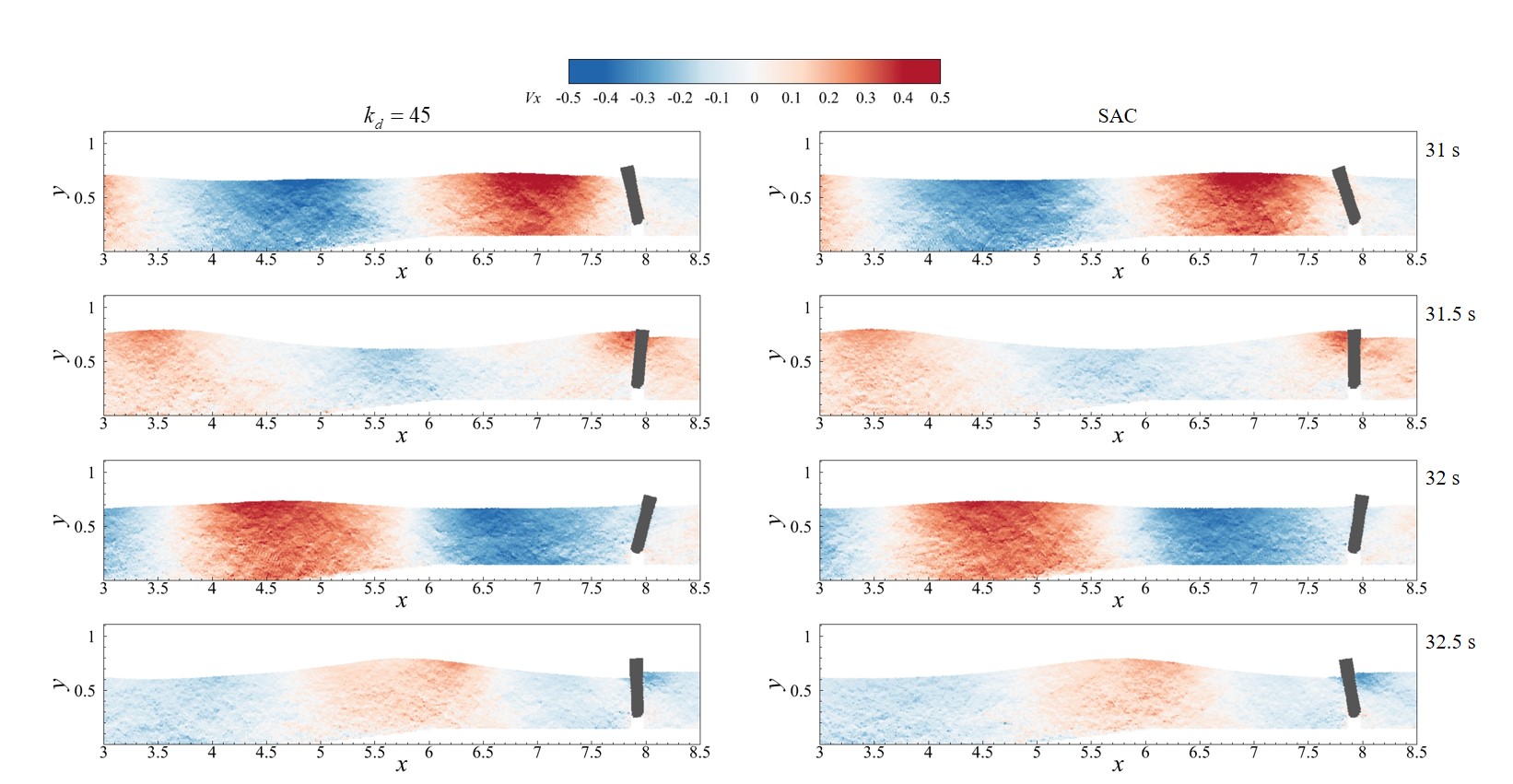}
\caption{Free surfaces and wave-structure interactions in one wave period. The fluid particles are colored by velocity magnitude on the x-axis.}\label{fig14}
\end{figure}

\begin{figure}[htbp]
    \centering
    \begin{subfigure}{\textwidth}
        \centering
        \includegraphics[width=\textwidth]{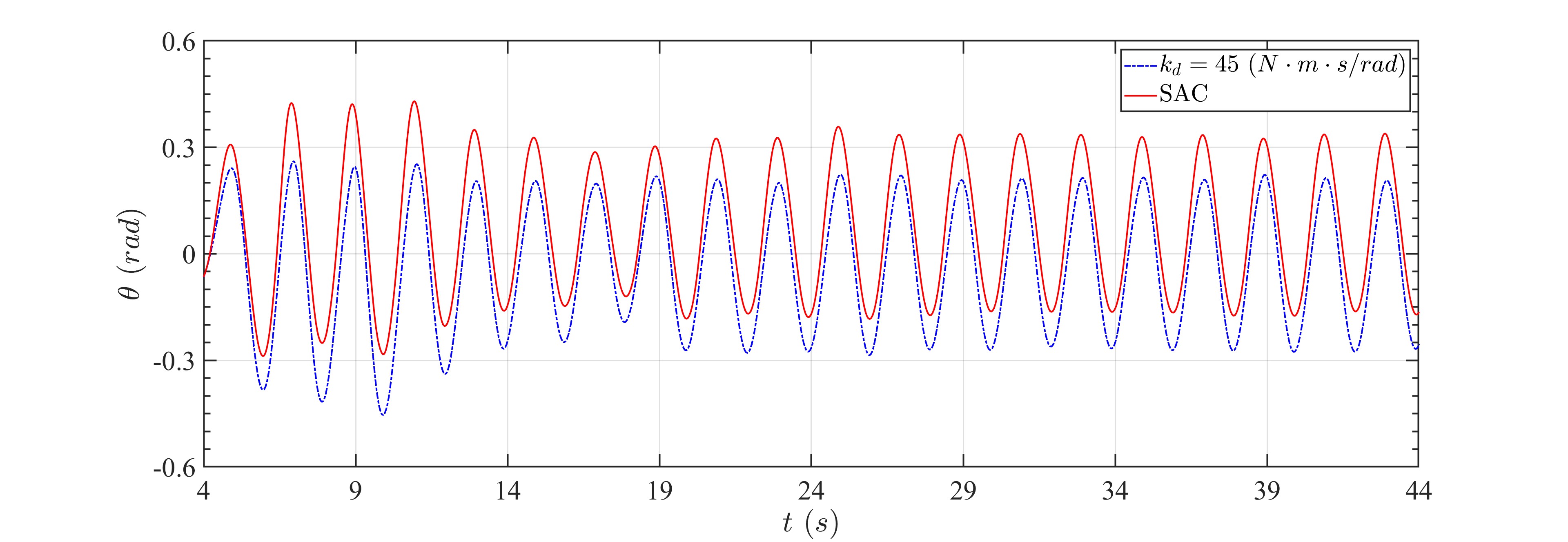}
        \caption{Rotation of the flap}
        \label{fig15a}
    \end{subfigure}
    
    \begin{subfigure}{\textwidth}
        \centering
        \includegraphics[width=\textwidth]{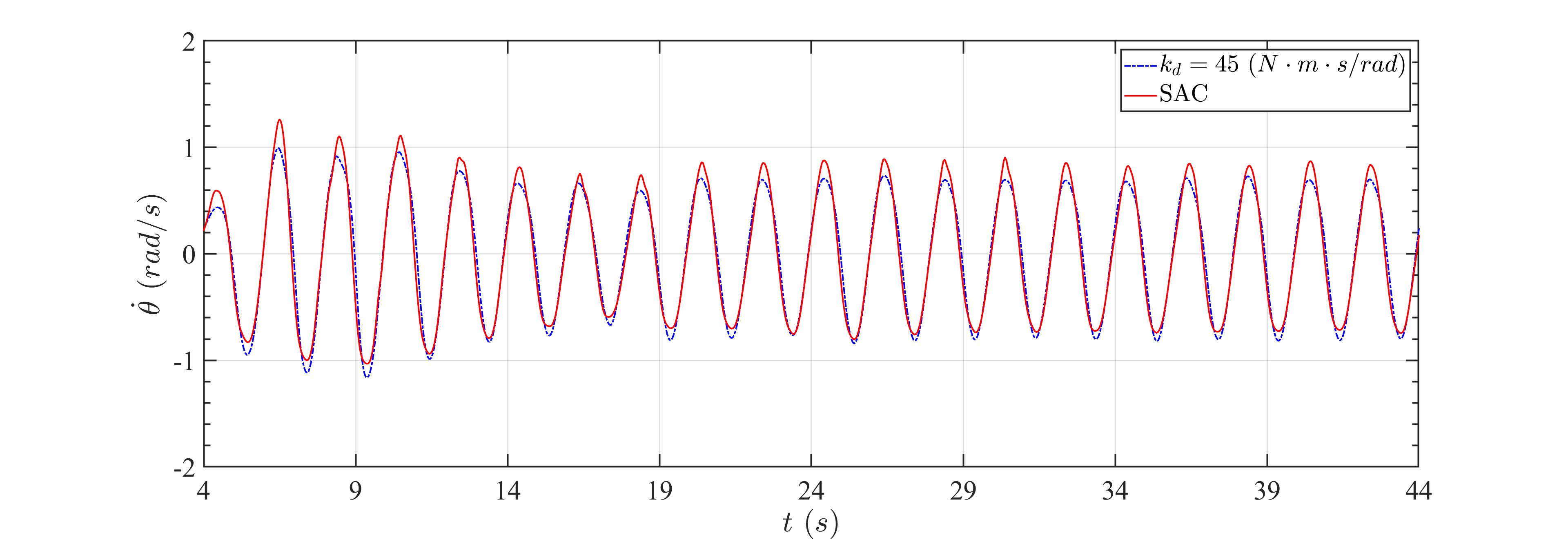}
        \caption{Angular velocity of the flap}
        \label{fig15b}
    \end{subfigure}
    
    \begin{subfigure}{\textwidth}
        \centering
        \includegraphics[width=\textwidth]{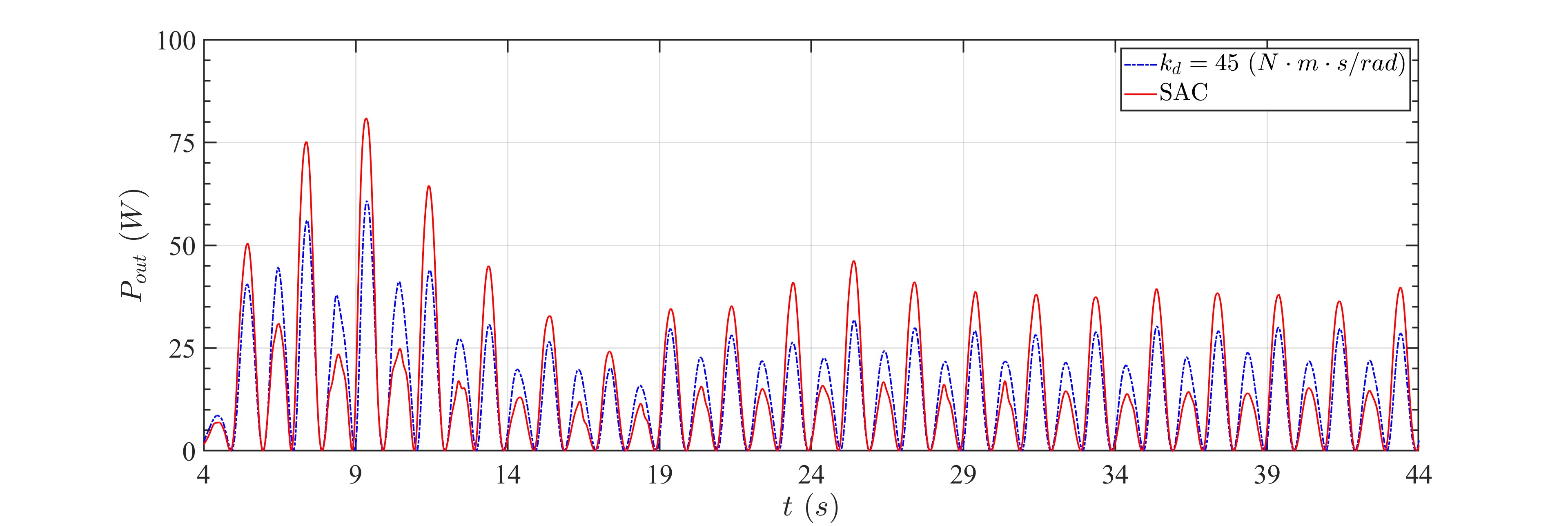}
        \caption{Instantaneous power capture}
        \label{fig15c}
    \end{subfigure}
    
    \caption{Influence of the damping coefficient on the (a) rotation of the flap, (b) angular velocity of the flap, and (c) instantaneous power capture.}
    \label{fig15}
\end{figure}

\begin{figure}[htbp]
    \centering
    \begin{minipage}[b]{0.45\textwidth}
        \centering
        \includegraphics[width=\textwidth]{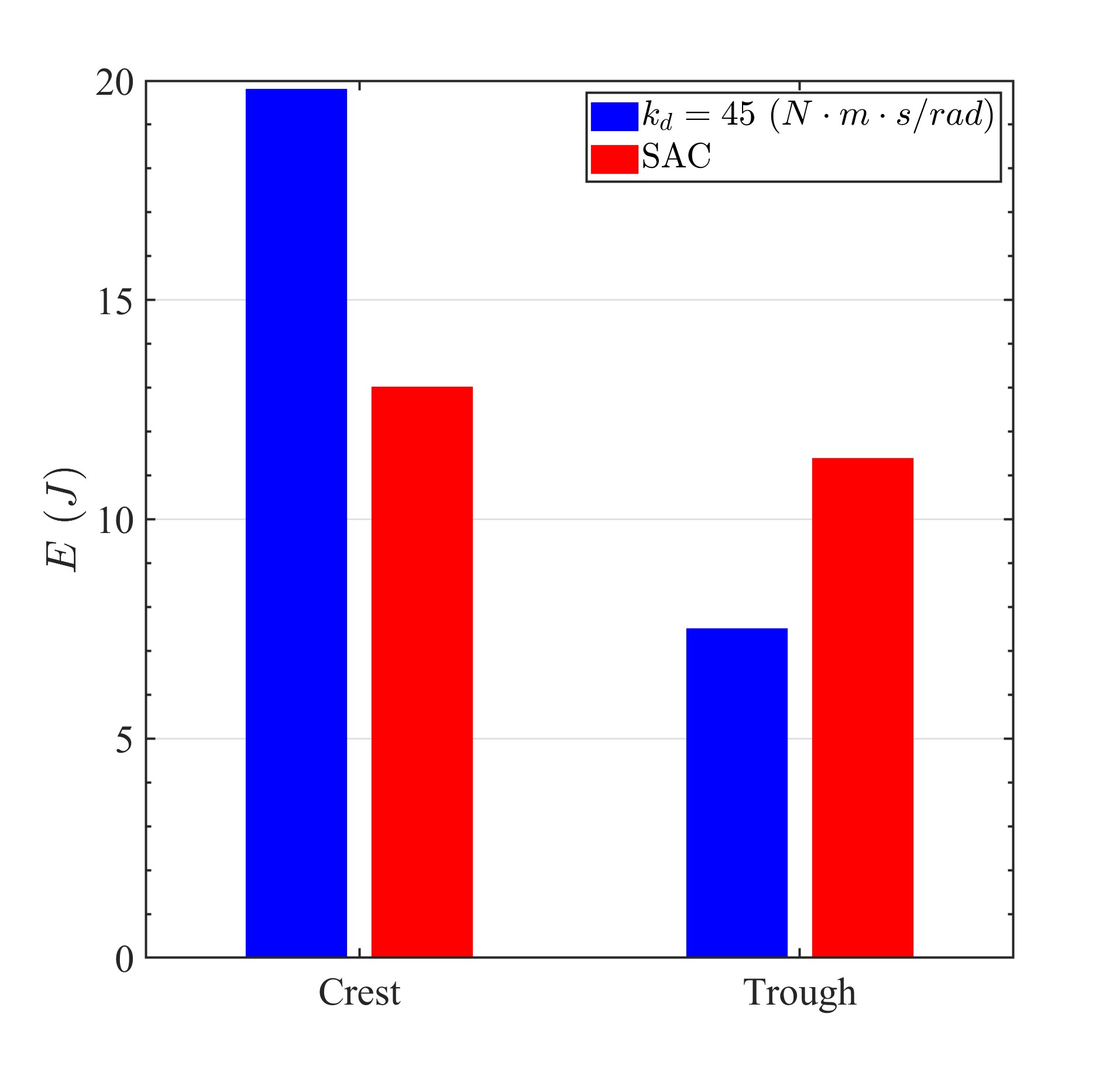}
        \subcaption{2D wave period}\label{fig16a}
    \end{minipage}
    \begin{minipage}[b]{0.45\textwidth}
        \centering
        \includegraphics[width=\textwidth]{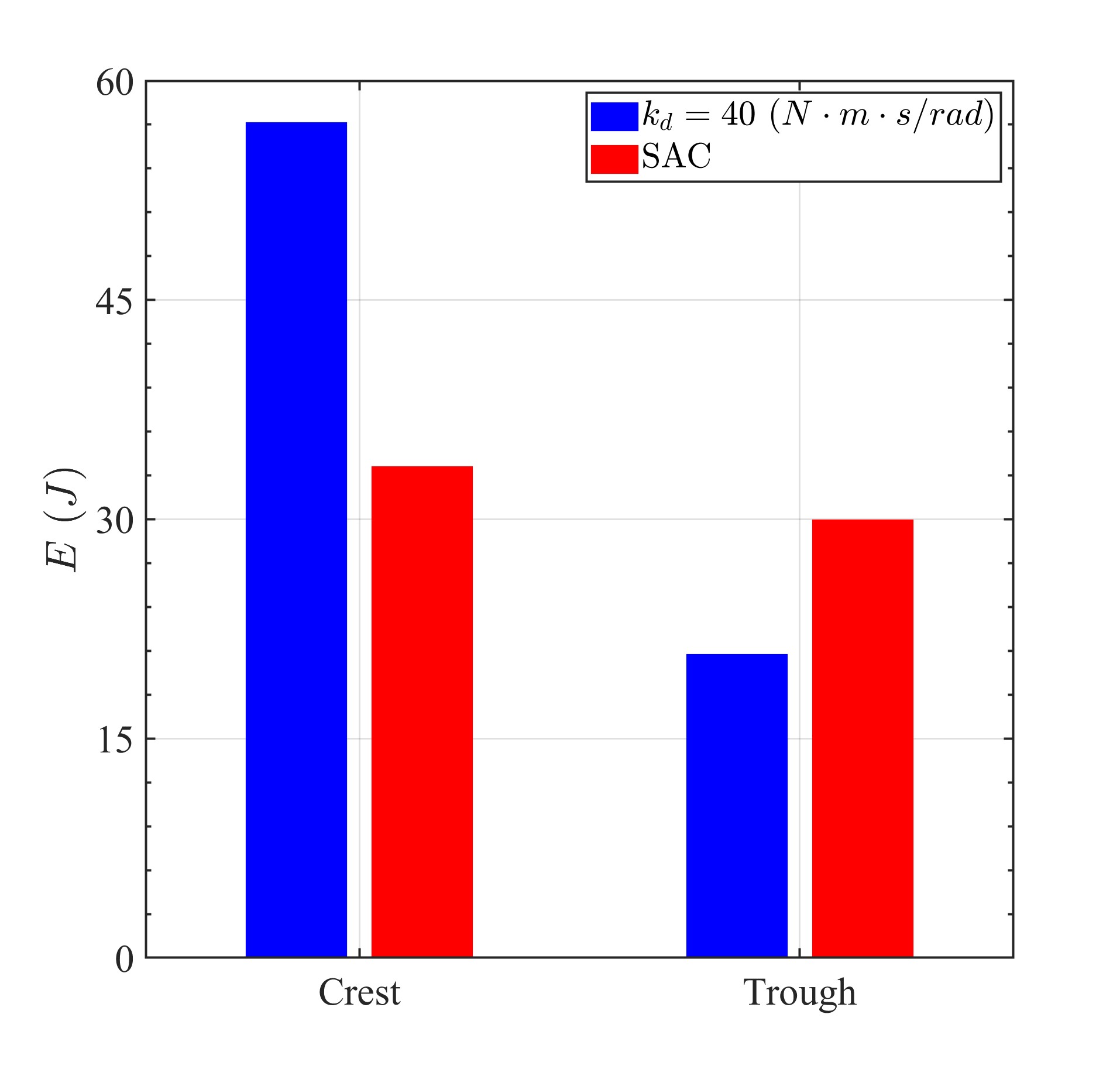}
        \subcaption{3D wave period}\label{fig16b}
    \end{minipage}
    \caption{Comparison of peak and trough energy in a complete cycle under the optimal constant damping coefficient and SAC strategy, (a) 2D and (b) 3D.}
    \label{fig16}
\end{figure}

Fig.\ref{fig14} presents the velocity field in the x-direction and the motion state of the flap under both fixed and dynamic damping coefficients, and Fig.\ref{fig15} focuses on quantitatively analyzing the flap's rotation and angular velocity. It can be observed that the adaptive change of the damping coefficient does not alter the flow field structure or the angular amplitude of the flap, while shift the equilibrium position of the flap, from -1.81° to 4.31°. Further analysis indicates that the periodic characteristics of the wave height at the flap are consistent with the damping coefficient of the PTO system. When the wave crest passes, the damping coefficient increases to its peak value. Given that the energy density of the wave crest is high, the angular velocity is reduced slightly, leading to an overall improvement in the PTO system, as shown in Fig.\ref{fig15} (c). In addition, the wave energy density of the trough itself is lower than that of the peak, and some energy has already been absorbed during the crest phase. Maintaining a high damping coefficient during this phase would rapidly decrease angular velocity. Although reducing the damping coefficient can increase the flap's angular velocity, the PTO system's power output still decreases compared to a constant damping coefficient. Overall, as shown in Fig.\ref{fig16} (b), during a complete wave period, the average energy harvesting by the dynamic damping system is $27.3 \ \text{J}$, compared to $24.4 \ \text{J}$ captured under the optimal constant damping coefficient, resulting in a 10.61\% improvement in wave energy harvesting.

The energy harvesting efficiency of the OWSC can be quantified by CWR \cite{brito2020numerical, senol2019enhancing}
\begin{equation}
\label{eq:eq42}
\text{CWR} = \frac{P_{out}}{P_0}.
\end{equation}
Here, $P_{out}$ is the capture of instantaneous energy within a wave period and $P_0$ is the mean incident power of unidirectional regular waves based on the linear theory
\begin{equation}
\left\{
\begin{aligned}
\label{eq:eq43}
P_{out} = \frac{1}{T_{period}} \int_{0}^{T_{period}} k_d \Omega^2 \, dt \\
P_0 = \frac{\rho g H^2 B \omega}{16 k} (1 + \frac{2kh}{\sinh(2kh)}),
\end{aligned}
\right.
\end{equation}
with $B$ denoting the width of the flap. Currently, CWR under optimal constant damping coefficient is 13.12\%, and 14.68\% for the adaptive damping coefficient in 2D simulations, as shown in Table~\ref{table3}.

\begin{table}[htbp]
    \centering
    \caption{Comparison of constant and adaptive damping coefficients for different wave types}
    \label{table3}
    \begin{tabular}{@{}lcc@{}}
        \toprule
          & Fixed $k_d$ (\%) & DRL (\%) \\ \midrule
        2D Regular Wave & 13.12 & 14.68 \\ 
        3D Regular Wave & 34.15 & 41.86 \\ 
        2D Irregular Wave & 86.79 & 92.38 \\ \bottomrule
    \end{tabular}
\end{table}

\subsection{Effects of 3D simulations}
Previous studies have shown that fixed flaps in 2D simulations simplify the diffraction waves, which are theoretically equal in size to the incident waves and opposite in direction. This results in standing waves forming on the windward side of the flap \cite{renzi2014wave}. In 3D simulations, diffraction waves propagate in all directions, and antisymmetric shear waves along the flap can trigger near-resonance, enhancing the torque acting on the converter. Therefore, 3D simulations can more accurately capture the motion characteristics of the OWSC device in actual operation. In this section, we conduct experiments in 3D environment with the policy obtained from 2D training environment. Notably, Zhang et al. \cite{zhang2021efficient} have proved that in 3D simulations, $k_d = 40$ N $\cdot$ m $\cdot$ s / rad is the optimal constant damping coefficient.

\begin{figure}[htbp]
\centering
\includegraphics[width=\textwidth]{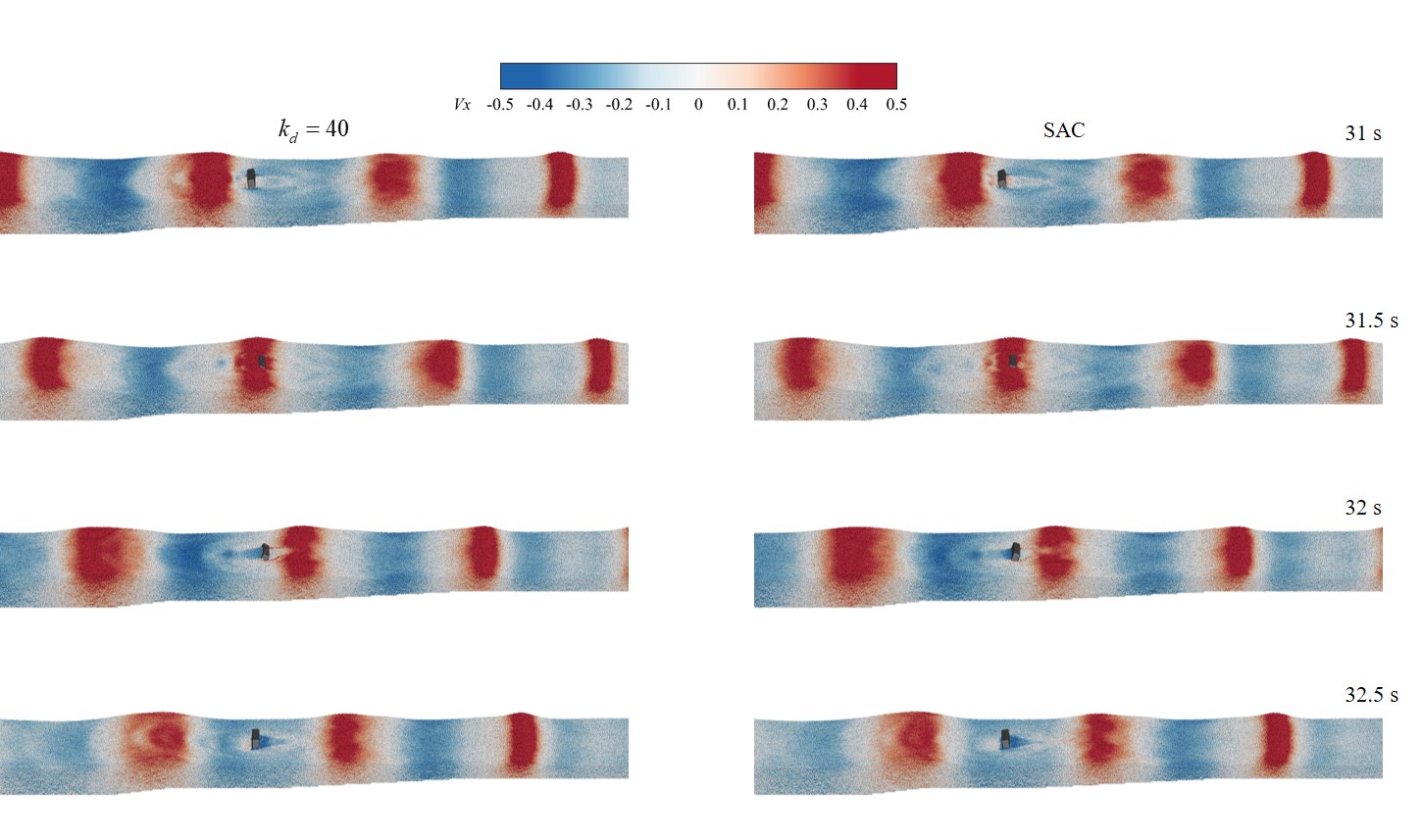}
\caption{3D simulations of OWSC: Free surfaces and the flap motion in a wave period. left side is the optimal constant damping coefficient and right side is the damping coefficient controlled with SAC. }\label{fig17}
\end{figure}

\begin{figure}[htbp]
    \centering
    \includegraphics[width=\textwidth]{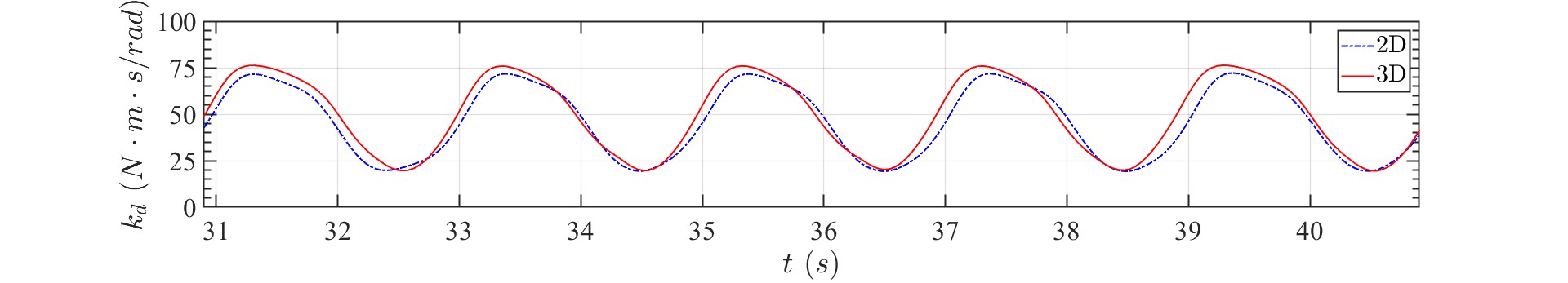}
    \caption{Comparison of the damping coefficient with the same policy under 2D and 3D simulations.}
    \label{fig18}
\end{figure}

As shown in Fig.~\ref{fig17}, the dynamic changes in the damping coefficient do not alter the structure of the flow field in the 3D simulations. Also as shown in Fig.~\ref{fig18}, the output of the damping coefficient in both 2D and 3D simulations is essentially consistent. This consistency indicates that the 2D assumption can accurately represent the coupling effect between the waves and the OWSC, and it also demonstrates the robustness of the trained policy network, which can be applied in real-world scenarios.

\begin{figure}[htbp]
    \centering
    \begin{subfigure}{\textwidth}
        \centering
        \includegraphics[width=\textwidth]{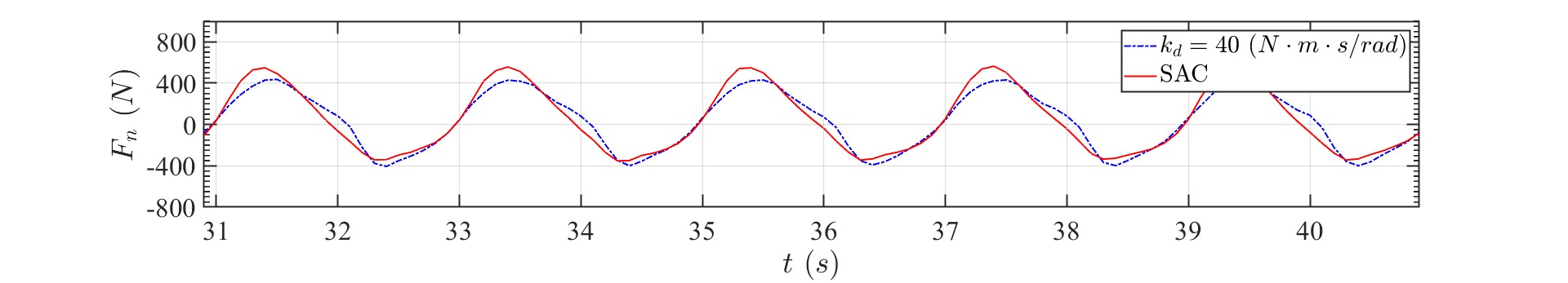}
        \caption{Vertical force on the flap}
        \label{fig19a}
    \end{subfigure}
    
    \begin{subfigure}{\textwidth}
        \centering
        \includegraphics[width=\textwidth]{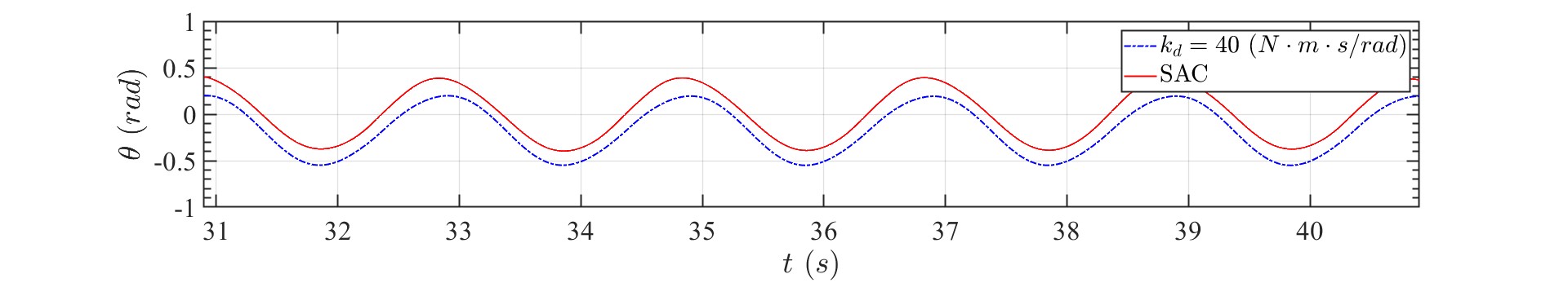}
        \caption{Rotation of the flap}
        \label{fig19b}
    \end{subfigure}
    
    \begin{subfigure}{\textwidth}
        \centering
        \includegraphics[width=\textwidth]{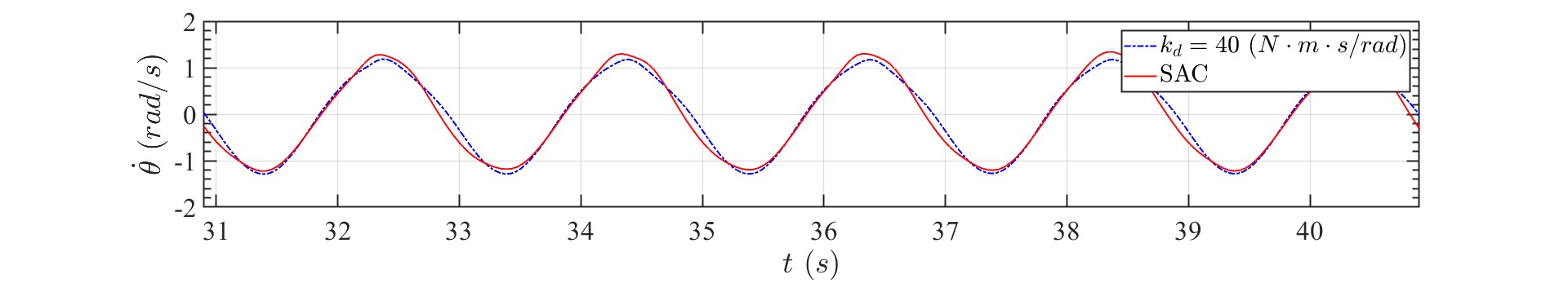}
        \caption{Angular velocity of the flap}
        \label{fig19c}
    \end{subfigure}
    
    \begin{subfigure}{\textwidth}
        \centering
        \includegraphics[width=\textwidth]{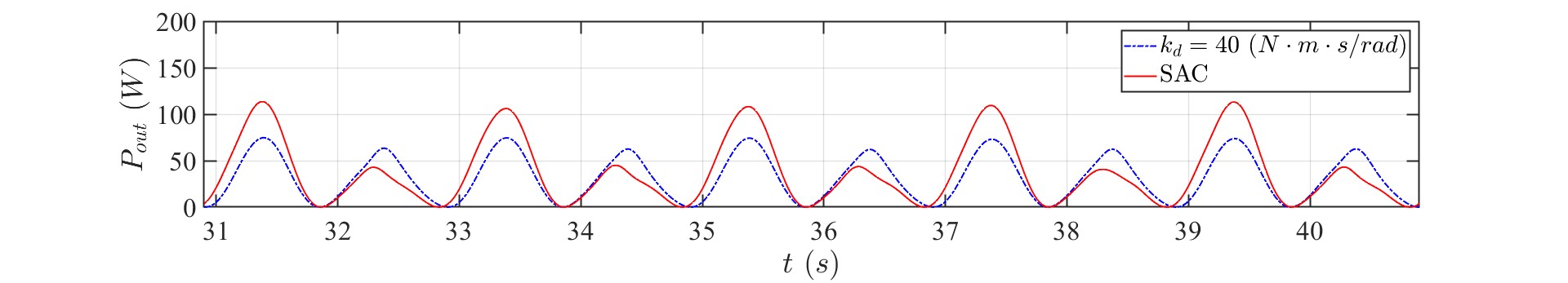}
        \caption{Instantaneous power capture}
        \label{fig19d}
    \end{subfigure}
    
    \caption{Effects of different policies on (a) vertical force on the flap, (b) rotation of the flap, (c) angular velocity of the flap, and (d) instantaneous power capture.}
    \label{fig19}
\end{figure}

Similar to the 2D simulation, Fig.~\ref{fig19} shows that under adaptive damping control, the equilibrium position of the flap shifts to the left by 11.46°, approaching nearly vertical position to the base. Considering the fluid incompressibility, this shift increases the force perpendicular to the flap, as illustrated in Fig.~\ref{fig19} (a). A control period is assumed to begin when the flap rotates to the far left, with the wave crest reaching the flap. During the first quarter of the period, the damping coefficient continues to rise. Due to the increased thrust on the flap, the angular velocity remains almost unchanged compared to the constant damping coefficient, enhancing energy harvesting. As the wave trough passes, the force on the flap decreases, and the reduction in the damping coefficient helps maintain the flap's angular velocity. Since the energy of the wave crest is inherently higher than that of the trough, the overall energy harvesting improves due to the difference in energy levels, as shown in Fig.~\ref{fig16} (b). Over a complete wave period, the average energy harvesting by the dynamic damping system is $77.88 \ \text{J}$, compared to $63.55 \ \text{J}$ under the optimal constant damping coefficient, resulting in a 22.54\% improvement. Also, in 3D simulation, CWR under optimal constant damping coefficient is 34.15\% and 41.86\% with the adaptive damping coefficient.

\subsection{Optimization under irregular waves}
In this section, we investigate optimization problems under irregular wave conditions, the most common scenarios encountered in practical engineering applications. Considering the strong nonlinearity of the induced motion of the OWSC under irregular waves, the number of observation points increases. The initial observation position of the wave height is set at $x = 3.5 \ \text{m}$, with one point placed every 0.264 m for total 17 probes. The locations of observation points for wave speed are also increased accordingly, and the value of the observation vector increased to 74 at last. In addition, to verify that the policy network can resist the random characteristics of irregular waves, two sets of random seeds are used to characterize the wave phases in Eq.~(\ref{eq:eq23}) during the training and testing stages while $H_{p}$ and $T_{p}$ remained unchanged. The relation between the total energy conversion and linear damping coefficient for irregular waves is shown in Table~\ref{table4}. It is clear that with $k_d = 60$ N $\cdot$ m $\cdot$ s / rad, the energy harvesting factor is the highest.

\begin{table}[htbp]
    \centering
    \caption{The variations of the total energy conversion in terms of damping coefficients.}
    \label{table4}
    \resizebox{\textwidth}{!}{
    \begin{tabular}{ccccccccccc}
        \toprule
        \(k_d\) (N $\cdot$ m $\cdot$ s / rad) & 10 & 20 & 30 & 40 & 50 & 60 & 70 & 80 & 90 & 100 \\ \midrule
        \(E_{train}\) (J) & 195.21 & 290.79 & 340.34 & 365.88 & 378.75 & 383.36 & 382.02 & 379.29 & 377.94 & 368.95 \\
        \(E_{test}\) (J) & 206.12 & 300.77 & 350.17 & 378.48 & 383.89 & 390.49 & 381.86 & 376.41 & 368.35 & 363.64 \\ \bottomrule
    \end{tabular}
    }
\end{table}

\begin{figure}[htbp]
    \centering
    \begin{subfigure}{\textwidth}
        \centering
        \includegraphics[width=\textwidth]{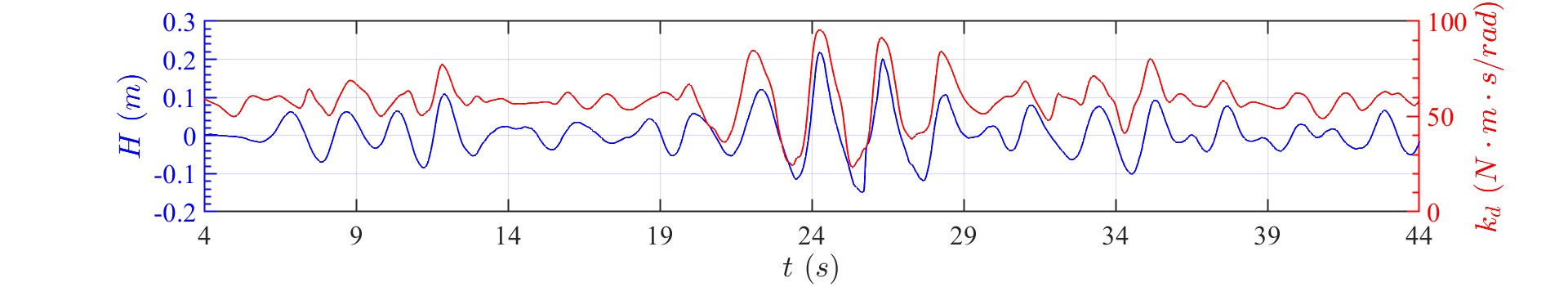}
        \caption{Training}
        \label{fig20a}
    \end{subfigure}

    \begin{subfigure}{\textwidth}
        \centering
        \includegraphics[width=\textwidth]{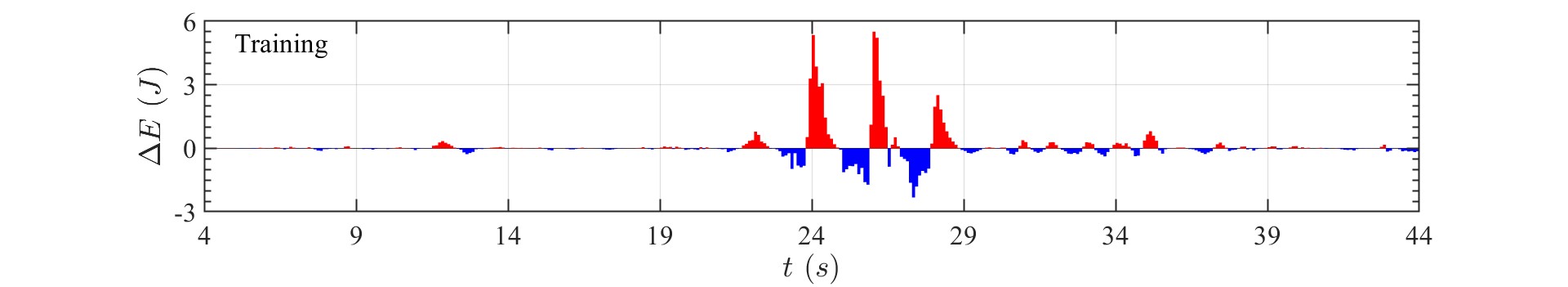}
        \caption{Training}
        \label{fig20b}
    \end{subfigure}

    \begin{subfigure}{\textwidth}
        \centering
        \includegraphics[width=\textwidth]{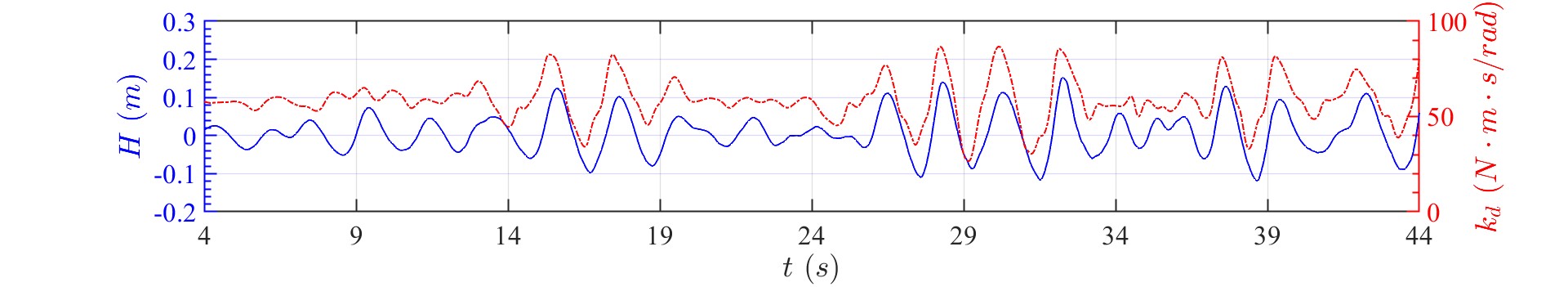}
        \caption{Testing}
        \label{fig20c}
    \end{subfigure}

    \begin{subfigure}{\textwidth}
        \centering
        \includegraphics[width=\textwidth]{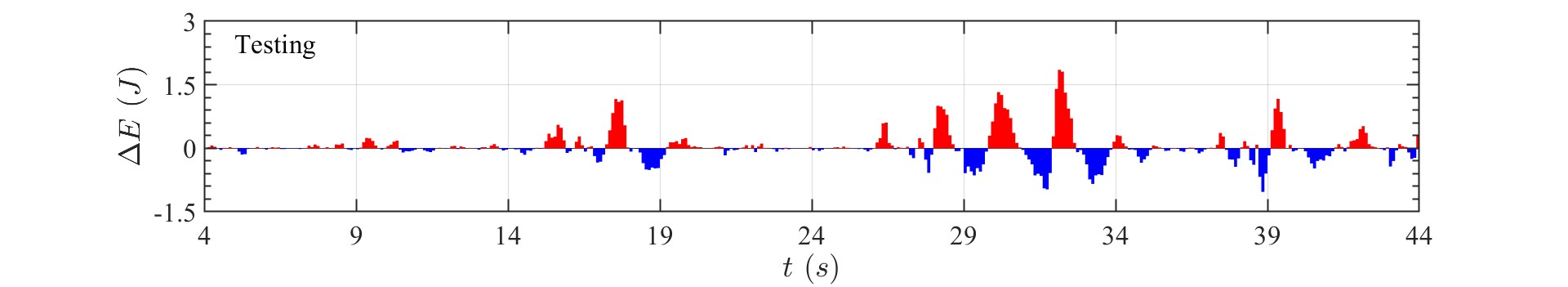}
        \caption{Testing}
        \label{fig20d}
    \end{subfigure}
    
    \caption{The wave height curve in front of the flap and the corresponding damping coefficient under the (a) training wave (c) testing wave, and the difference of instantaneous energy harvesting $\Delta E = E_t - E_{60}$ under the (a) training wave (c) testing wave.}
    \label{fig20}
\end{figure}

From Fig.~\ref{fig20} (a), it can be observed that, compared to regular waves, less than one-third of the wave heights of irregular waves exceed 0.2 m throughout the entire period, which is also the region where the wave energy is primarily concentrated. The dynamic response of the damping coefficient is related to the wave height. When the peak period occurs, the damped vibration also increases accordingly. This relationship can also be observed in the test section, indicating that the agent can accurately capture the wave characteristics under the specific spectrum.

Further combined with Fig.~\ref{fig20} (b), we can see that compared with the constant damping coefficient, the difference in energy harvesting is mainly concentrated in the peak period, which is essentially consistent with the improvement of energy harvesting by regular waves. The near-simple harmonic damping motion will improve Capture energy in the crest section and reduce energy harvesting in the trough section. For secondary period waves, since the instantaneous energy they carry is small, the response of the damping coefficient will not cause significant changes in flap motion and energy harvesting. This part of the energy cannot be effectively improved. Therefore, over the entire 40-second period, the wave energy harvesting increased by 24.67 J, an increase of 6.42\%, compared to the energy harvesting under a constant damping coefficient. In addition, for the average incident energy of the irregular wave, Eq.~(\ref{eq:eq43}) is modified as
\begin{equation}
\label{eq:eq44}
P_0 = \sum_{i=0}^{N} \frac{\rho g H_i^2 B \omega_i}{16 k_i} \left(1 + \frac{2k_i h}{\sinh(2k_i h)}\right).
\end{equation}
Then, CWR under optimal constant damping coefficient is 86.79\% and 92.38\% for the adaptive damping coefficient.

\subsection{Study of dual OWSC system}
\begin{figure}[htbp]
\centering
\includegraphics[width=\textwidth]{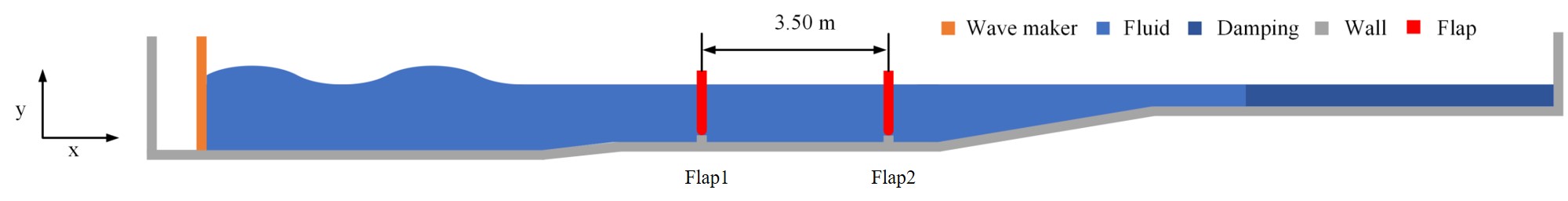}
\caption{The geometry of the dual OWSC system. The overall structure has stayed the same. Only the length of the plane where the base is located has been increased.}\label{fig21}
\end{figure}

The dual OWSC system is illustrated in Fig.~\ref{fig21}. Previous research has indicated that for a dual OWSC system, the maximum total energy conversion is achieved when the spacing between the two OWSCs is seven-eighths of the wavelength \cite{chow2018experimental}. In this section, we initially set the damping coefficients of both flaps to 50 N $\cdot$ m $\cdot$ s / rad to investigate the impact of different spacings on total energy conversion. As shown in Table~\ref{table5}, when the spacing is 3.5 m, approximately three-quarters of the wavelength, the total energy conversion reaches its maximum.

\begin{table}[htbp]
    \centering
    \caption{The influence of the spacing on the total energy conversion of dual OWSCs.}
    \label{table5}
    \resizebox{\textwidth}{!}{
    \begin{tabular}{cccccccccc}
        \toprule
        \(\Delta x\) (m) & 2.0 & 2.25 & 2.5 & 2.75 & 3.0 & 3.25 & 3.5 & 3.75 & 4.0 \\ \midrule
        \(E_t\) (J) & 449.01 & 366.32 & 422.91 & 526.69 & 608.52 & 661.68 & 675.47 & 650.44 & 541.84 \\ \bottomrule
    \end{tabular}
    }
\end{table}

\begin{figure}[htbp]
\centering
\includegraphics[width=0.6\textwidth]{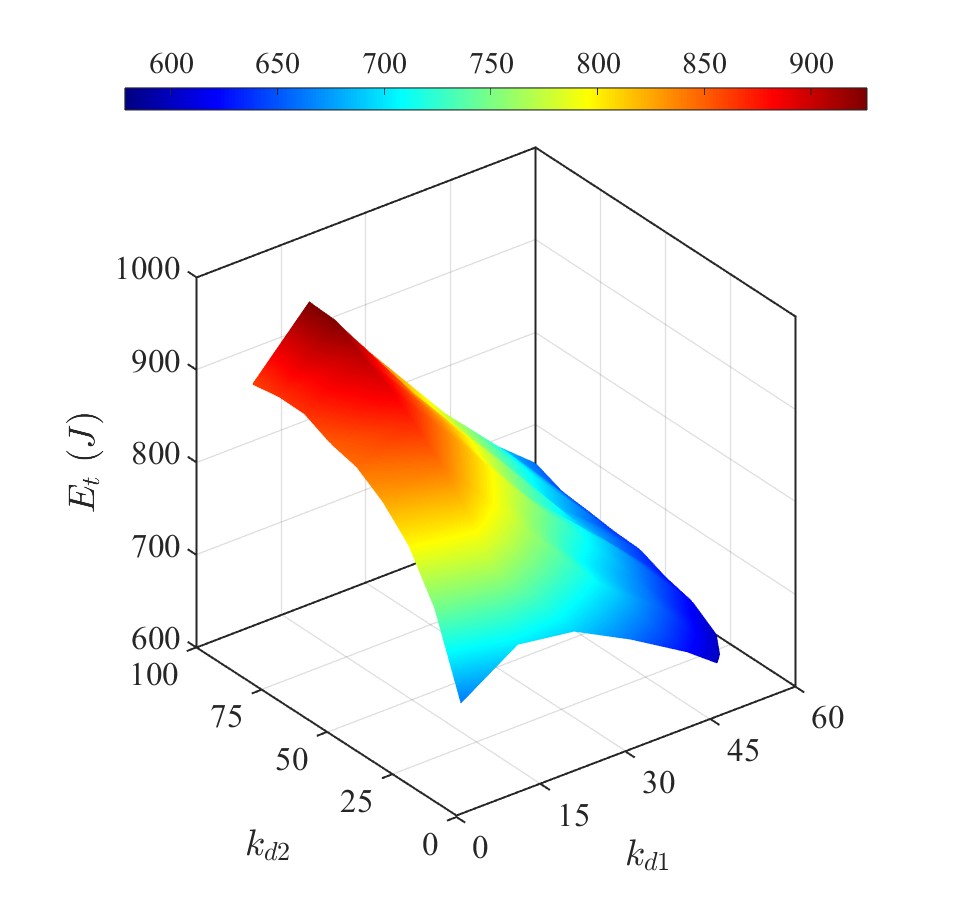}
\caption{The variations of the total energy conversion in terms of the damping coefficients in dual OWSCs.}\label{fig22}
\end{figure}

Subsequently, we examine the effect of different damping coefficient combinations on total energy harvesting at this optimal spacing, as depicted in Fig.~\ref{fig22}. It can be observed that the damping coefficient of flap-2 has a linear relationship with the total energy conversion, where increasing the damping coefficient results in a gradual increase in energy harvesting. On the other hand, for flap-1, total energy conversion increases initially with the damping coefficient and then decreases. Notably, the peak value occurs around $k_{d1} = 20$ N $\cdot$ m $\cdot$ s / rad, which remains unaffected by changes in $k_{d2}$. Therefore, in this section, the condition with $k_{d1} = 20$ N $\cdot$ m $\cdot$ s / rad and $k_{d2} = 80$ N $\cdot$ m $\cdot$ s / rad will be used as the baseline for RL training. In addition, considering that the dual OWSC system has strong nonlinear characteristics and flap-2 has low energy harvesting, in the training of this section, the damping coefficient of OWSC-2 is kept unchanged and only the damping coefficient of OWSC-1 is controlled. The training starts from the 24th second, when the coupling effect of the wave structure tends to be stable.

Subsequently, the effect of varying damping coefficient combinations on the total energy harvesting at the identified optimal spacing is investigated, as depicted in Fig.~\ref{fig22}. The analysis reveals a linear relationship between the damping coefficient of the OWSC-2 and the total energy conversion, with higher damping coefficients leading to a gradual increase in energy harvesting. Conversely, for the OWSC-1, the total energy conversion initially increases with the damping coefficient, reaching a peak before declining. Notably, the peak value occurs around $k_{d1} = 20$ N $\cdot$ m $\cdot$ s / rad, and this peak remains unaffected by variations in $k_{d2}$. Therefore, for the subsequent RL training, the condition with $k_{d1} = 20$ N $\cdot$ m $\cdot$ s / rad and $k_{d2} = 80$ N $\cdot$ m $\cdot$ s / rad is established as the baseline.
Given the strong nonlinear characteristics of the dual OWSC system and the observed lower wave energy harvesting by the OWSC-2, the damping coefficient of the OWSC-2 is held constant during training. This approach allows us to focus on optimizing the damping coefficient of the OWSC-1. The RL training commences at the 24th second, a point in time when the wave-structure interactions have stabilized, meaning the system has reached a steady state in terms of energy harvesting and conversion.

\begin{figure}[htbp]
\centering
\includegraphics[width=\textwidth]{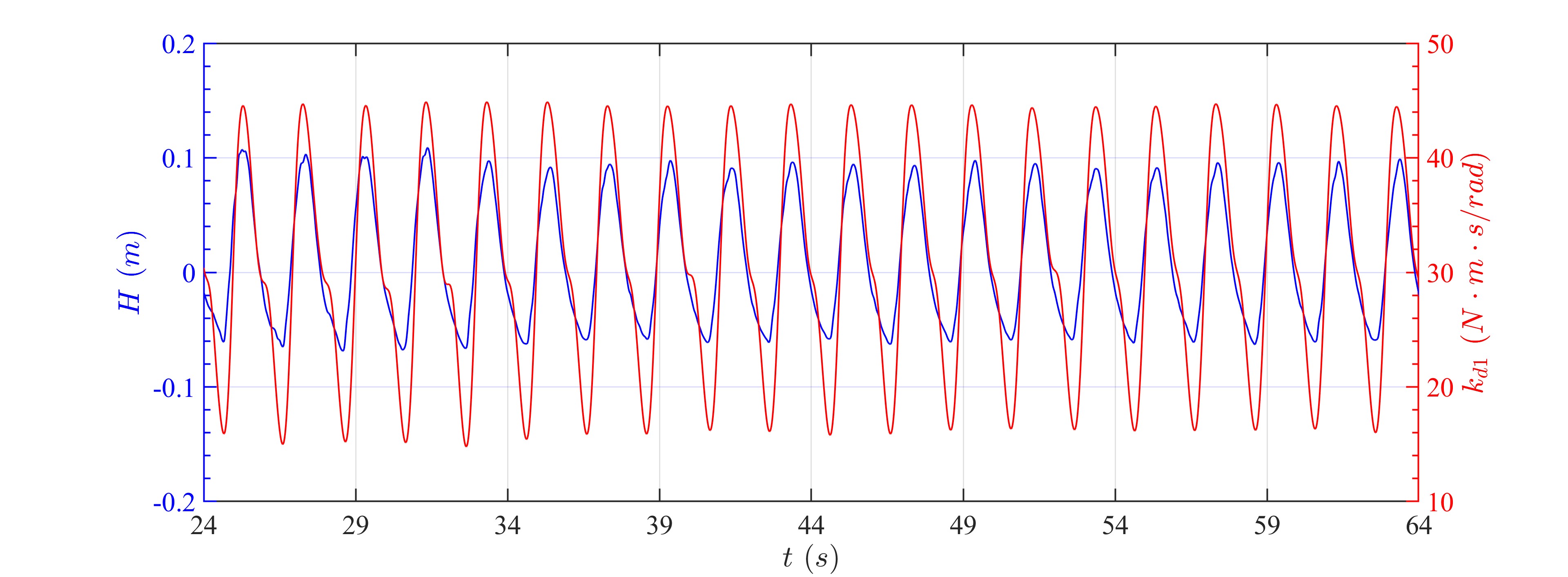}
\caption{The wave height in front of the first flap ($x = 7.6$ $m$) and the corresponding damping coefficient.}\label{fig23}
\end{figure}

\begin{figure}[htbp]
    \centering
    \begin{subfigure}{\textwidth}
        \centering
        \includegraphics[width=\textwidth]{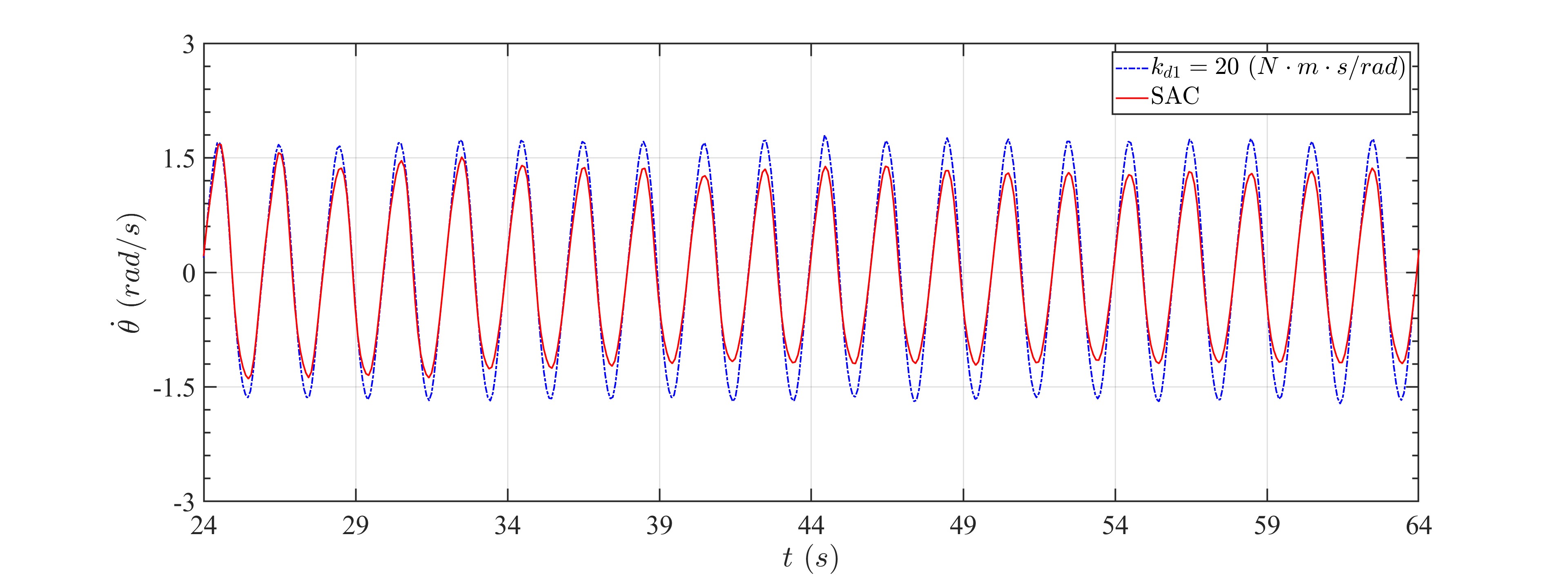}
        \caption{Angular velocity of the flap}
        \label{fig24a}
    \end{subfigure}

    \begin{subfigure}{\textwidth}
        \centering
        \includegraphics[width=\textwidth]{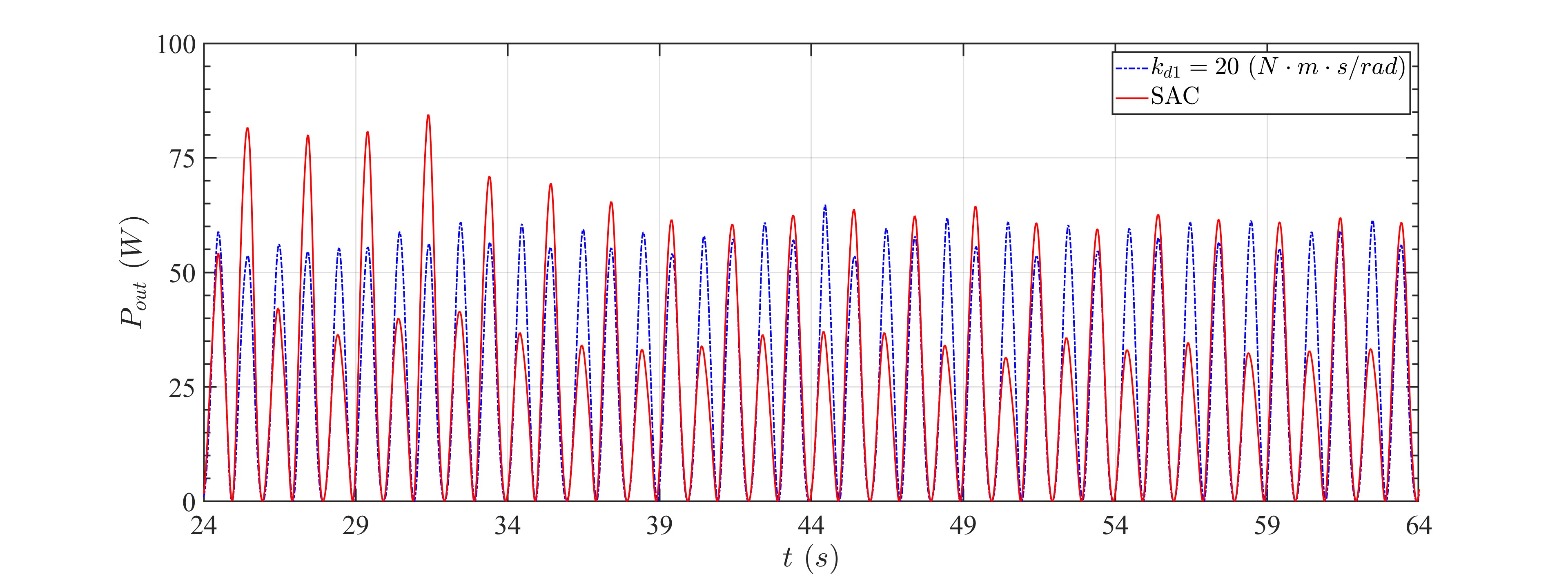}
        \caption{Instantaneous power capture}
        \label{fig24b}
    \end{subfigure}
    
    \caption{The influence of the damping coefficient on the (a) angular velocity of the first flap, and (b) instantaneous power capture.}
    \label{fig24}
\end{figure}

The training results are illustrated in Fig.~\ref{fig23}. The damping coefficient variation is consistent with the trend of wave height changes before the flap, demonstrating apparent periodicity that aligns with previous research findings. Further analysis based on Fig.~\ref{fig24} reveals that, after 39 seconds, the system's state is stabilized. An increase in the damping coefficient during the wave peak phase significantly reduces the angular velocity, resulting in only a limited increase in energy harvesting during the peak. Conversely, reducing the damping during the trough phase does not enhance the angular velocity, which remains lower than under constant damping conditions. This reduction in energy harvesting during the trough phase leads to a notable decrease in overall energy acquisition over the entire wave period, thereby failing to improve energy harvesting efficiency. During the 60-second test period, the energy harvesting for the OWSC-1 using the adaptive damping coefficient was 1560.37J, compared to 1626.52J with constant damping, representing a reduction of 4.07\%.

\begin{figure}[htbp]
\centering
\includegraphics[width=\textwidth]{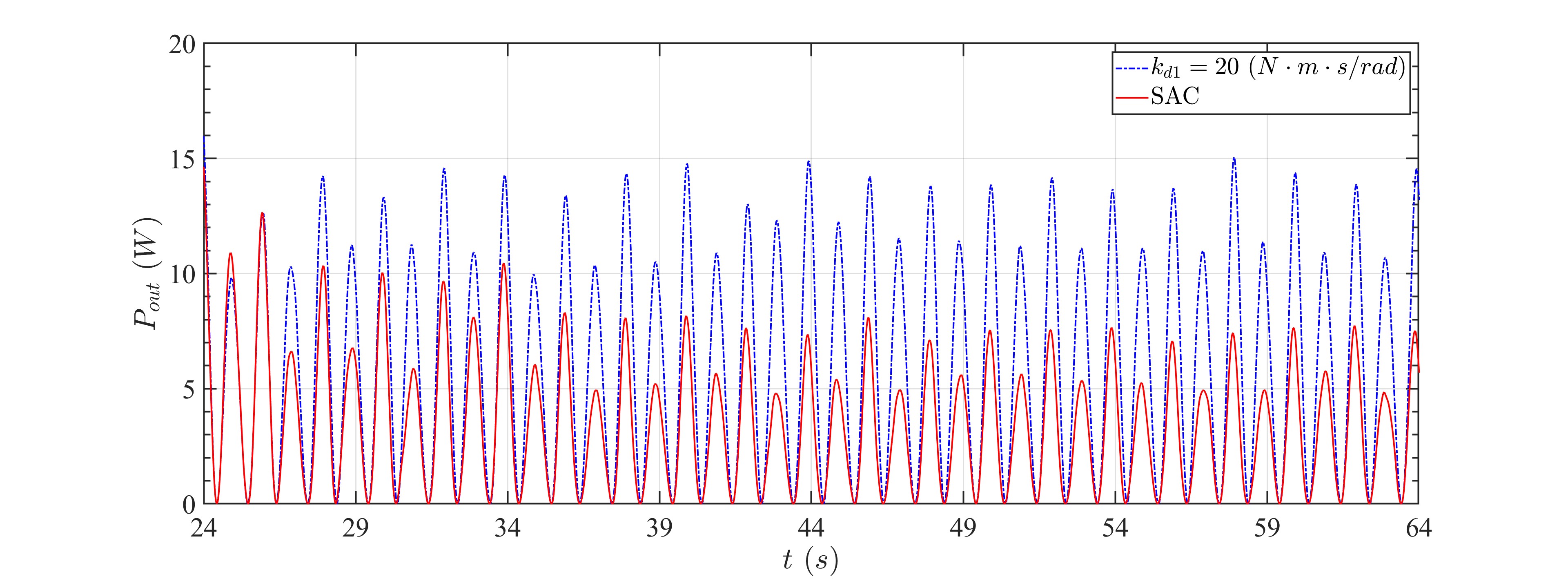}
\caption{The influence of the damping coefficient on the instantaneous power capture of the OWSC-2.}\label{fig25}
\end{figure}

\begin{figure}[htbp]
\centering
\includegraphics[width=\textwidth]{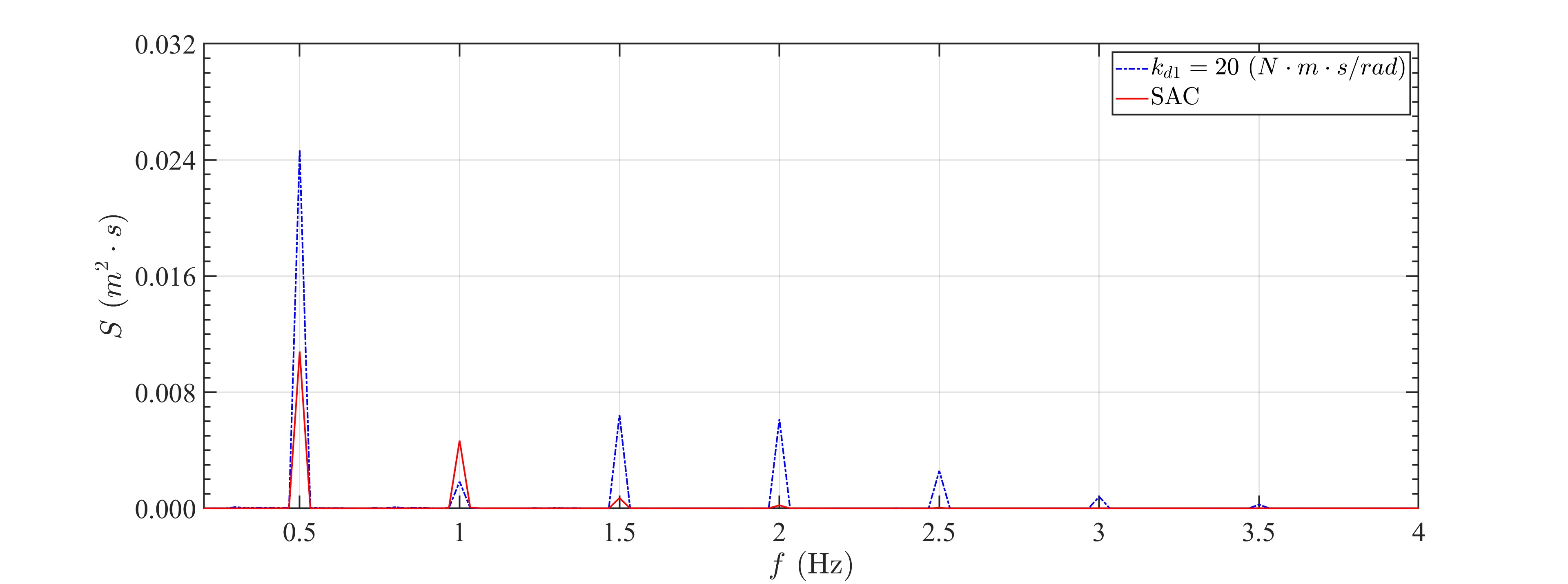}
\caption{The spectrum diagram at $x = 8.2$ $m$.}\label{fig26}
\end{figure}

Additionally, for the OWSC-2, where the damping coefficient remained unchanged, the energy harvesting efficiency dropped significantly from 362.94 J to 197.91 J, a decrease of 45.47\%, as shown in Fig.~\ref{fig25}. This indicates that after the wave passes through the OWSC-1 with adaptive damping, the energy loss is more significant than with constant damping. Combined with Fig.~\ref{fig26}, it is evident that under constant damping, significant harmonics are generated between the two OWSCs, which is beneficial for enhancing energy harvesting.

Therefore, in the dual OWSC system, the nonlinear characteristics are pronounced, and single damping control is insufficient to improve overall wave energy acquisition efficiency. Moreover, the 2D simulations constrain the design and optimization of the OWSC layout, necessitating further analysis and discussion in subsequent work.

\section{Conclusion}
This paper establishes a framework coupling a CFD environment based on an open-source SPH-based library with DRL, aimed at optimizing the adaptive damping coefficient of the PTO system in OWSCs for wave energy conversion. Initially, the wave-making model and the numerical model of WSI were validated. Subsequently, the performance of various RL algorithms for the optimization process was investigated. The results indicated that SAC considers policy entropy, balanced exploration, and exploitation well and provides effective policies to enhance wave energy conversion.

For regular waves, the strategy primarily utilizes the difference in energy density between wave crests and troughs. Increasing energy harvesting during the crest phase and reducing it during the trough phase achieves a positive net energy harvesting over each wave period. The policy trained in 2D simulations can be effectively applied in 3D simulations. Although the 2D simulations simplify wave diffraction and result in a slight decrease in calculated CWR, they accurately capture the coupling characteristics between waves and OWSCs, allowing DRL to learn practical policies that are robust and transferable. This provides a theoretical foundation for validation in experiments.

The DRL algorithm could still learn effective energy conversion optimization policies for irregular waves, primarily targeting regular waves with high energy density in the main period. The optimization principle is similar to that for regular waves, with limited enhancement in energy harvesting from the dynamic damping response for waves in the secondary period due to their lower energy density.

Finally, this paper explores the optimization of wave energy absorption in a dual OWSC system. The interaction between incident waves and OWSCs in the dual system generates harmonics with strong nonlinearity. Optimization focused on the primary OWSC showed that using similar optimization policies can not enhance energy harvesting, as it significantly reduces the wave energy density between the OWSCs, leading to a substantial decrease in energy absorption by the secondary OWSC and an overall reduction in the system's energy harvesting. Therefore, considering the dynamic response of a single OWSC's damping coefficeint is insufficient to optimize the energy conversion in complex dual OWSC systems. 

Future work will introduce multi-agent reinforcement learning to directly learn corresponding energy optimization strategies for various OWSC layouts in 3D simulations.

\section*{CRediT authorship contribution statement}
\textbf{M. Ye:} Validation, Software, Methodology, Investigation, Formal analysis, Writing - original draft, Writing - review \& editing, Visualization. \textbf{C. Zhang:} Software, Writing – review \& editing, Investigation. \textbf{Y.R. Ren:} Resources, Formal analysis, Conceptualization. \textbf{Z.Y. Liu:} Methodology, Data curation. \textbf{O.J. Haidn:} Validation, Supervision, Investigation. \textbf{X.Y. Hu:} Writing – review \& editing, Supervision, Methodology, Investigation, Conceptualization.

\section*{Declaration of competing interest}
The authors declare that they have no known competing financial interests or personal relationships that could have appeared to influence the work reported in this paper.

\section*{Data availability}
The corresponding code of this work is available on GitHub at \url{https://github.com/Xiangyu-Hu/SPHinXsys}. The corresponding data of this work will be made available on reasonable request.

\section*{Acknowledgments}
M. Ye was supported by China Scholarship Council (No.202006120018) when he conducted this work.

\bibliographystyle{elsarticle-num} 
\bibliography{owsc}

\end{document}